\documentclass[aps,prd,twocolumn,showpacs]{revtex4}
\usepackage{graphics}
\usepackage{amssymb}
\usepackage{amsmath}
\usepackage{epsfig}
\usepackage{slashed}
\usepackage{color}
\usepackage{verbatim}
\usepackage{float}
\usepackage{mathrsfs}
\usepackage{esint}

\begin{document}
\title{Fermi-liquid theory of imbalanced quark matter}

\author{J.J.R.M. van Heugten}
\author{Shaoyu Yin\footnote{s.yin@uu.nl}}
\author{H.T.C. Stoof}
\affiliation{Institute for Theoretical Physics, Utrecht University, Leuvenlaan 4, 3584 CE Utrecht, The Netherlands}

\begin{abstract}
The temperature dependence of the thermodynamic potential of quantum chromodynamics (QCD), the specific heat, and the quark effective mass are calculated for imbalanced quark matter in the limit of a large number of quark flavors (large-$N_F$), which corresponds to the random phase approximation. Also a generalization of the relativistic Landau effective-mass relation in the imbalanced case is given, which is then applied to this thermodynamic potential.
\end{abstract}
\maketitle

\section{Introduction}

Landau Fermi-liquid theory has seen considerable success in describing a wide variety of fermionic many-particle systems, such as liquid Helium-3, electrons in metals, nuclei and nuclear matter \cite{Nozieres:1997,Negele:1998,Baym:2004,Mahan:2000}. It gives an effective description of the low-lying elementary excitations (quasiparticles) at low temperatures, i.e., at a temperature $T$ such that the system can be considered degenerate ($T\ll\mu$, where $\mu$ denotes the chemical potential) but still in the normal phase above any symmetry breaking phase transition ($T\gg T_c$), for example, to a magnetic or superconducting phase. Nevertheless, the theory is not only important for the description of the above ``normal'' (Fermi-liquid) phase, but is also vital to correctly describe the emergence of a possible ordered phase \cite{Leggett:1999}. Indeed, according to the Bardeen-Cooper-Schrieffer (BCS) theory the onset of superconductivity is to be viewed as an instability of the Fermi liquid under an attractive interaction, which results in the formation of Cooper pairs \cite{Schrieffer:1999}. The same is true for magnetism in the case of repulsive interactions.

In particular, BCS theory implies that the high-density and low-temperature region of the phase diagram of quantum chromodynamics (QCD) contains a color superconducting phase of quarks \cite{Alford:2008,Bailin:1984}. In fact, since quarks carry color, flavor and spin quantum numbers, many distinct superconducting phases are possible and are characterized by the various symmetries of the Cooper-pair wavefunction. These phases of cold and dense QCD might occur in the core of neutron stars where matter is compressed to several times the nuclear density $\rho_0\simeq0.16\,\mathrm{fm}^{-3}$.
The presence of such a superconducting phase is expected to have observable consequences on the cooling and magnetic fields of neutron stars \cite{Becker:2008,Haensel:2006,Blaschke:2006}. A particularly interesting phase is the so-called 2SC phase in which only two flavors of quarks are paired, while the unpaired quarks of the third flavor remain a Fermi liquid. This phase further illustrates the importance of understanding the quark-gluon plasma, i.e., the normal phase of quarks.

Fortunately, due to asymptotic freedom, the coupling constant $g$ of QCD becomes small at large chemical potential $\mu$, such that a systematic study of the high-density and low-temperature region of the QCD phase diagram is possible using perturbation theory. At high densities the dominant interaction between quarks is that of one-gluon exchange, where the long-range behavior of the gluons is screened due to the quark-gluon plasma. While the electric gluons (longitudinal) are screened by a Debye mass $m_g\sim g\mu$, the magnetic (transverse) gluons are only dynamically screened \cite{Le Bellac:1996,Kapusta:2006}. This residual long-range behavior of the magnetic gluons dominates the low-temperature behavior of the system. Examples of the effects of magnetic gluons can be seen in the non-standard scaling of the critical temperature of the superconducting phase $T_c\sim\mu g^{-5}\exp\left(-3\pi^2/g\right)$ \cite{Son:1999,Pisarski:2000,Hong:2000}, which depends exponentially only on $1/g$ instead of showing the expected $1/g^2$ behavior, or of the life-time of elementary excitations of (grand-canonical) energy $E$ near the Fermi surface $\tau(E)\sim|E|^{-1}$ \cite{Manuel:2000,Le Bellac:1997,Schafer:2004,Brown:2000} that signal a (marginal) non-Fermi-liquid behavior at zero temperature.

In the context of a possible deconfined phase of quarks inside neutron stars, the Fermi system is expected to be imbalanced due to the different chemical potentials of the various quarks. This deconfined phase may contain $u$, $d$, and $s$ quarks with different densities as a consequence of their different masses and charges. In cold atomic physics, an analogous population imbalance has been realized experimentally between two spin states \cite{Zwierlein:2006,Partridge:2006}, which resulted in frustration of the pairing between particles. This will cause very different behavior from the normal BCS case and is currently a very hot topic in a wide variety of fields \cite{Casalbuoni:2004,Bedaque:2003}. In particular, the imbalance between different flavors of quarks is expected to significantly alter the properties of the QCD phase diagram, such as the 2SC phase mentioned above or the color-flavor locked phase where all the colors and flavors are paired \cite{Alford:2008}. In the present paper, we therefore generalize the quasiparticle properties based on Fermi-liquid theory to an imbalanced system.

To do so, the effect of the dressed gluons on the thermodynamic potential, specific heat and effective mass of the quasiparticles is determined at low temperatures. The thermodynamic potential will be calculated for a two-flavor quark system in the limit of a large number of quark flavors (large-$N_F$). This approximation corresponds to the random phase approximation (RPA), which has been quite successful in condensed-matter systems such as the interacting electron gas in metals \cite{Rice:1965}. Using the framework of Landau Fermi-liquid theory for relativistic systems, derived in section \ref{sec-theory}, the effective mass of the quasiparticles can be determined from the thermodynamic potential by considering the specific heat. The result and its implications on the applicability of Fermi-liquid theory is discussed. As we will show, the logarithmic dependence of the temperature, which is due to the transverse gluons, shows the breakdown of Fermi-liquid theory at low temperatures. Furthermore, even though within our approximations there is no interaction between the different flavors of quarks in the limit of weak coupling, the RPA correction still presents some mixing between the different flavors, which stems from the long-wavelength screening of the longitudinal gluons. Above and throughout the article natural units are used, i.e., units such that $\hbar=c=k_B=1$. Other conventions and technical details of the calculation can be found in the Appendices.

\section{Landau Fermi-liquid theory}\label{sec-theory}

To describe the normal state of the quark-gluon plasma at nonzero
temperatures, which is a strongly interacting gas of quarks, Landau
Fermi-liquid theory can be used. This theory takes as a starting point
the non-interacting Fermi gas and switches on the interaction adiabatically.
As long as the temperature of the system is much higher than the critical
temperature $T_{c}$ for superconductivity, no bound states (Cooper
pairs) will form and each state of the non-interacting Fermi gas is
transformed into a state of the interacting gas. Therefore, the excitations
of such a Fermi liquid remain of a fermionic nature. For simplicity,
Fermi-liquid theory will be introduced here for a system at zero temperature
even though the quark-gluon plasma is formally a marginal Fermi liquid
in that case. However, the introduced concepts turn out to be valid
for nonzero temperatures much lower than the Fermi temperature $T_F$,
which is expected to be around $10^{13}~\mathrm{K}$ for quarks in the core of neutron stars.

The basis of Landau Fermi-liquid theory is to consider the effect
of a small change of the ground-state Fermi distribution on the thermodynamic
potential density, which can be written as
\begin{align}\label{thermpotential}
\delta\Omega=&-\sum_\sigma n_\sigma\delta\mu_\sigma+\frac{1}{V}\sum_{\mathbf{k},\sigma}(\epsilon_\sigma(\mathbf{k})-\mu_\sigma)\delta N_{\sigma}(\mathbf{k})\nonumber\\
&+\frac{1}{2V^2}\sum_{\mathbf{k},\sigma;\mathbf{k}',\sigma'}f_{\sigma\sigma'}(\mathbf{k},\mathbf{k}')\delta N_\sigma(\mathbf{k})\delta N_{\sigma'}(\mathbf{k}').
\end{align}
Here $\delta N_\sigma(\mathbf{k})$ is a small change in the ground-state momentum
distribution of species $\sigma$, $n_\sigma$ is the particle density, $\epsilon_\sigma(\mathbf{k})$ is the energy of a quasiparticle that for our purposes only depends on the magnitude of the vector $\mathbf{k}$, and $\mu_\sigma$ is the chemical potential. In the case of electrons the index $\sigma$ specifies the spin of the electron, while for the case of quarks it specifies the spin, color and flavor of the quark. Hence, the quasiparticle energy and the effective interaction between quasiparticles are defined as
\begin{align*}
\epsilon_\sigma(\mathbf{k})-\mu_\sigma&\equiv\frac{\delta\Omega}{\delta N_\sigma(\mathbf{k})},\\
f_{\sigma\sigma'}(\mathbf{k},\mathbf{k}')&\equiv\frac{\delta\Omega}{\delta N_\sigma(\mathbf{k})\delta N_{\sigma'}(\mathbf{k}')}.
\end{align*}

An important quantity in Fermi-liquid theory is the effective mass $m_\sigma^*$ of a quasiparticle of species $\sigma$ near its Fermi surface. It is defined in terms of the group velocity $v_\sigma^*$ evaluated at the Fermi momentum $k_\sigma$
\begin{equation}
m_{\sigma}^*\equiv\frac{k_\sigma}{v_\sigma^*}\equiv k_\sigma\left[\frac{\partial\epsilon_{\sigma}(\mathbf{k})}{\partial k}\right]_{k=k_\sigma}^{-1},
\end{equation}
such that upon linearizing the quasiparticle energy around the Fermi surface we obtain
\begin{equation}
\epsilon_{\sigma}(\mathbf{k})\simeq\mu_\sigma+\frac{k_\sigma}{m_\sigma^*}(k-k_\sigma),
\end{equation}
with $\mu_\sigma=\epsilon_\sigma(k_\sigma)$ as the Fermi energy of species $\sigma$. In the non-interacting limit, the effective mass reduces to $m_\sigma^*=m_\sigma$ and $m_\sigma^*=\epsilon_{\sigma0}(k_\sigma)=\sqrt{k_\sigma^2+m_\sigma^2}$
for non-relativistic and relativistic dispersions, respectively.

\begin{figure}
\begin{centering}
\includegraphics[scale=0.25]{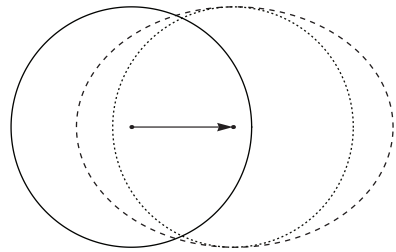}
\par\end{centering}
\caption{A Galilean transformation (dotted curve) and Lorentz boost (dashed curve) of a Fermi sphere (solid curve). The boosts have been scaled in order to compare the shapes of the Fermi sphere after the transformation. The volume of the Fermi sphere, i.e., the density of the gas, becomes larger due to the Lorentz contraction.\label{fig:Boosted_Fermi_Sphere}}
\end{figure}

Using the fact that the pressure is an invariant under Galilean or Lorentz transformations (see Appendix.~\ref{app-invariance}), of which the effect can be written as a change in the distribution as shown in Fig.~\ref{fig:Boosted_Fermi_Sphere}, Eq.~(\ref{thermpotential}) should give zero for this particular choice of $\delta N_{\sigma}(\mathbf{k})$ and hence relate $\epsilon_\sigma(\mathbf{k})$ to $f_{\sigma\sigma'}(\mathbf{k},\mathbf{k}')$. In other words, by performing an infinitesimal Galilean or Lorentz transformation on the Fermi sphere, the effective mass can be related to the effective interaction at the Fermi surface \cite{Nozieres:1997,Baym:1976}.

To this end, consider an infinitesimal change in the Fermi surface
from $k_\sigma$ to $k_\sigma+\delta k_\sigma(\hat{\mathbf{k}})$
\[
\delta N_{\sigma}(\mathbf{k})=\begin{cases}
1 & k_\sigma\leq k\leq k_\sigma+\delta k_\sigma(\hat{\mathbf{k}})\quad\delta k_\sigma>0,\\
-1 & k_\sigma+\delta k_\sigma(\hat{\mathbf{k}})\leq k\leq k_\sigma\quad\delta k_\sigma<0,\\
0 & \textnormal{otherwise},
\end{cases}
\]
where the change in the Fermi surface solely depends on the direction in momentum space denoted by $\hat{\mathbf{k}}$. The corresponding change in the thermodynamic potential density can be written as
\begin{widetext}
\begin{align}\label{thermpotexpand}
\delta\Omega=&\sum_\sigma\int\frac{\mathrm{d}\hat{\mathbf{k}}}{(2\pi)^3}\intop_{k_\sigma}^{k_\sigma+\delta k_\sigma}\mathrm{d}kk^2[\epsilon_\sigma(\mathbf{k})-\mu_\sigma]+\frac{1}{2}\sum_{\sigma,\sigma'}\iint\frac{\mathrm{d}\hat{\mathbf{k}}}{(2\pi)^3}\frac{\mathrm{d}\hat{\mathbf{k}}'}{(2\pi)^3}\intop_{k_\sigma}^{k_\sigma+\delta k_\sigma}\mathrm{d}kk^2\intop_{k_{\sigma'}}^{k_{\sigma'}+\delta k_{\sigma'}}\mathrm{d}k'k'^2f_{\sigma\sigma'}(\mathbf{k},\mathbf{k}')-\sum_\sigma n_\sigma\delta\mu_\sigma\nonumber\\
=&\sum_{\sigma}\frac{k_\sigma^3}{4\pi^2m_\sigma^*}\int\frac{\mathrm{d}\hat{\mathbf{k}}}{4\pi}\delta k_\sigma^2(\hat{\mathbf{k}})+\sum_{\sigma,\sigma'}\frac{k_\sigma^2k_{\sigma'}^2}{2(2\pi^2)^2}\iint\frac{\mathrm{d}\hat{\mathbf{k}}}{4\pi}\frac{\mathrm{d}\hat{\mathbf{k}}'}{4\pi}f_{\sigma\sigma'}(k_\sigma,k_{\sigma'},\theta)\delta k_\sigma(\hat{\mathbf{k}})\delta k_{\sigma'}(\hat{\mathbf{k}}')-\sum_\sigma n_\sigma\delta\mu_\sigma.
\end{align}
\end{widetext}
In the second line the integrals were expanded to second order in $\delta k_\sigma(\hat{\mathbf{k}})$, and $f_{\sigma\sigma'}(k_\sigma,k_{\sigma'},\theta)$ was defined as the effective interaction at the Fermi surfaces between two species, with $\theta$ the angle between the directions of $\mathbf{k}$ and $\mathbf{k}'$. Next we expand $\delta k_\sigma$ in terms of spherical harmonics and $f_{\sigma\sigma'}$ in terms of Legendre polynomials as
\begin{align*}
\delta k_\sigma(\hat{\mathbf{k}})&=\sum_{l,m}\delta k_\sigma^{lm}Y_{lm}(\hat{\mathbf{k}}),\\
f_{\sigma\sigma'}(k_\sigma,k_{\sigma'},\theta)&=\sum_lf_{l,\sigma\sigma'}(k_\sigma,k_{\sigma'})P_l(\cos\theta).
\end{align*}
For an imbalanced system, the difference in density $n_\sigma$ results in a difference in $\mu_\sigma$, which can cause $m^*_\sigma$ to be different from each other even if their non-interacting bare masses $m_\sigma$ are the same. For each species, the particle density obeys $n_\sigma=k_\sigma^3/6\pi^2$. Inserting these expansions and using the orthogonality of the spherical harmonics, Eq.~(\ref{thermpotexpand}) becomes
\begin{align}\label{reducedthermpotexpansion}
\delta\Omega=&-\sum_\sigma n_\sigma\delta\mu_\sigma+\sum_{l,m,\sigma}\frac{3n_\sigma|\delta k_\sigma^{lm}|^2}{2m_\sigma^*}\nonumber\\
&+\sum_{l,m,\sigma,\sigma'}\frac{f_{l,\sigma\sigma'}(k_\sigma,k_{\sigma'})k_\sigma^2k_{\sigma'}^2\delta k_\sigma^{lm}(\delta k_{\sigma'}^{lm})^*}{8\pi^4(2l+1)}.
\end{align}
In the balanced case, all Fermi momenta are the same, $k_\sigma=k_F$, $\delta k_\sigma^{lm}=\delta k_F^{lm}$, then Eq.~(\ref{reducedthermpotexpansion}) reduces to
\begin{equation}\label{LandauFermithermpot}
\delta\Omega=-n\delta\mu+\sum_{l,m}\frac{3n}{2m^*}|\delta k_F^{lm}|^2\left[1+\frac{F_l}{2l+1}\right],
\end{equation}
where the intensive quantities $m^*$, $\mu$, $k_F$ and $F_l$ are the same for all species, and $n$ is the total density, i.e., the density summed over all species. The (dimensionless) Landau Fermi-liquid parameters $F_l$ are defined as
\begin{equation}\label{LandauFermiparameter}
F_l=\sum_{\sigma'}\frac{k_F^2f_{l,\sigma\sigma'}(k_F)}{2\pi^2}\left[\frac{\partial\epsilon(\mathbf{k})}{\partial k}\right]_{k_F}^{-1}=\sum_{\sigma'}\frac{k_Fm^*f_{l,\sigma\sigma'}(k_F)}{2\pi^2}.
\end{equation}
Note that $F_l$ is species independent, but the inter-species and intra-species interactions can be different, so the subscript $\sigma$ is shown explicitly in $f_{l,\sigma\sigma'}$ but not in $F_l$. From the above it is possible to derive the equation for the effective mass, as we first show for the simpler balanced case.

Consider a change in the distribution due to a Lorentz transformation as shown in Fig.~\ref{fig:Boosted_Fermi_Sphere}. The shape of the Fermi sphere is determined from the equation $\mu=-k_\mu u^\mu$, where $k^\mu=(\epsilon(\mathbf{k}),\mathbf{k})$ is the four-momentum, $u^\mu=(\gamma,\gamma\mathbf{v})$ is the four-velocity of the Fermi sphere with the Lorentz factor $\gamma=1/\sqrt{1-v^2}$. The minus sign in the previous expression and in $u_\mu u^\mu=-1$ are a consequence of our choice of the metric $\eta_{\mu\nu}=\mathrm{diag}(-1,1,1,1)$. In the rest frame, where $u^\mu=(1,\mathbf{0})$, the above reduces to the condition $\epsilon(k_F)=\mu$. Similarly, the shape of the Fermi surface can be determined from the Lorentz transformation of the momentum to a frame moving with velocity $-\mathbf{v}$,
\[\mathbf{k}\rightarrow\mathbf{k}-\hat{\mathbf{v}}(\hat{\mathbf{v}}\cdot\mathbf{k})(1-\gamma)+\epsilon(\mathbf{k})\mathbf{v}\gamma,\]
which for an infinitesimal Lorentz transformation reduces to $\mathbf{k}\rightarrow\mathbf{k}+\epsilon(\mathbf{k})\mathbf{v}$
and in the non-relativistic limit to $\mathbf{k}\rightarrow\mathbf{k}+m\mathbf{v}$.
Note that the deformation of the Fermi sphere due to the Lorentz contraction
shows up only in higher-order terms of $\mathbf{v}$.

First, consider the non-relativistic case in which an infinitesimal
Galilean transformation shifts all momenta from $\mathbf{k}$ to $\mathbf{k}+m\mathbf{v}$. This results in a change of the Fermi surface according to $\delta k_F=m\hat{\mathbf{k}}\cdot\mathbf{v}=mvY_{10}(\cos\theta)/\sqrt{3}$,
such that $\delta k_F^{lm}=\delta_{l1}\delta_{m0}mv/\sqrt{3}$.
Furthermore, the total energy of the system transforms as $E\rightarrow E+\frac{1}{2}Mv^2$,
where $M$ is the total mass of the system, which induces a change
in the chemical potential $\mu=\partial E/\partial N\rightarrow\mu+\frac{1}{2}mv^2$.
Inserting this into Eq.~(\ref{LandauFermithermpot}) gives the well-known Landau effective-mass relation in the balanced case \cite{Nozieres:1997,Negele:1998,Mahan:2000,Baym:2004}
\begin{equation}
m^*=m\left(1+\frac{1}{3}F_1\right).\label{nonrel-Landaumass}
\end{equation}

Secondly, for a relativistic system the momenta are shifted by $\mathbf{k}\rightarrow\mathbf{k}+\epsilon(\mathbf{k})\mathbf{v}$
such that $\delta k_F^{lm}=\delta_{l1}\delta_{m0}\mu v/\sqrt{3}$.
Since the total energy of the system transforms as $E\rightarrow\gamma E\simeq E+\frac{1}{2}Ev^2$,
the change in the chemical potential is $\delta\mu=\frac{1}{2}\mu v^2$.
Inserting these into Eq.~(\ref{LandauFermithermpot}) gives the relativistic
Landau effective-mass relation \cite{Baym:1976}
\begin{equation}
m^*=\mu\left(1+\frac{1}{3}F_1\right).\label{rel-Landaumass}
\end{equation}
Note that in the right-hand side $F_1$ also depends on $\mu$, because the Fermi momentum $k_F$ is a, in general complicated, function of $\mu$.

Now for the more general imbalanced case. Since the Lorentz transformation is applied uniformly to each species, according to the symmetry of the formula, we can expect that the contribution from each species in Eq.~(\ref{reducedthermpotexpansion}) should vanish, namely,
\[\frac{n_\sigma\mu_\sigma^2v^2}{2m_\sigma^*}+\sum_{\sigma'}\frac{f_{1,\sigma\sigma'}(k_\sigma,k_{\sigma'})k_\sigma^2k_{\sigma'}^2\mu_\sigma\mu_{\sigma'}v^2}{72\pi^4}-\frac{n_\sigma\mu_\sigma v^2}{2}=0.\]
Then we get the general expression for the effective mass,
\begin{equation}\label{imbalancemass}
m_\sigma^*=\mu_\sigma\left[1+\sum_{\sigma'}\frac{f_{1,\sigma\sigma'}(k_\sigma,k_{\sigma'})N_{\sigma'}(0)}{3}\frac{\mu_{\sigma'}k_{\sigma'}m_\sigma^*}{\mu_\sigma k_\sigma m^*_{\sigma'}}\right],
\end{equation}
where $N_\sigma(0)=m_\sigma^*k_\sigma/2\pi^2$ is the density of states at the Fermi energy of species $\sigma$. The arguments in $f_1$ emphasize that, in the imbalanced case, the interaction may depend not only on the angle between the momenta but also on the absolute value of the Fermi momentum of each species. In fact, the same expression of $m_\sigma^*$ can be obtained more strictly by considering the addition of a single particle to the system and comparing the energy increase in two different frames.

We can introduce the average effective mass, which is of practical importance, as $m^*=\sum_\sigma m^*_\sigma/N$ with $N$ the total number of species. Correspondingly, the average chemical potential is $\mu=\sum_\sigma\mu_\sigma/N$. In order to get a similar expression as in Eq.~(\ref{rel-Landaumass}), we introduce the relative weight of each species as $x_\sigma=\mu_\sigma/N\mu$ such that $\sum_\sigma x_\sigma=1$, and then generalize the Landau effective-mass relation with average values as
\begin{equation}\label{averageimbalancemass} m^*=\mu\left[1+\frac{1}{3}\sum_{\sigma\sigma'}f_{1,\sigma\sigma'}(k_\sigma,k_{\sigma'})N_{\sigma'}(0)\frac{x_{\sigma'}k_{\sigma'}m_\sigma^*}{k_\sigma m^*_{\sigma'}}\right],
\end{equation}
where the double-sum term, which can be defined again as $F_1$, plays the role of the Landau parameter in the balanced case as in Eq.~(\ref{rel-Landaumass}), therefore we obtain now
\[F_1=\sum_{\sigma\sigma'}f_{1,\sigma\sigma'}(k_\sigma,k_{\sigma'})N_{\sigma'}(0)\frac{x_{\sigma'}k_{\sigma'}m_\sigma^*}{k_\sigma m^*_{\sigma'}},\]
which, obviously, reduces to Eq.~(\ref{LandauFermiparameter}) as the system becomes balanced.

In the following sections the effective mass will be derived from a microscopic RPA calculation. In particular, in Sec.~\ref{sec-thermpotential} it will be determined from the specific heat. The specific heat (per volume) for low temperatures can be expressed in terms of the effective masses by
\begin{align}\label{CV}
C_V&=\frac{1}{V}\left(\frac{\partial E}{\partial T}\right)_{V,N}=\frac{1}{V}\left[\sum_{k,\sigma}\frac{\partial E}{\partial N_\sigma(\mathbf{k})}\frac{\partial N_\sigma(\mathbf{k})}{\partial T}\right]_{V,N}\nonumber\\
&\simeq\sum_\sigma\frac{k_\sigma m^*_\sigma}{6}T=\sum_\sigma\frac{\pi^2}{3}N_\sigma(0)T,
\end{align}
where it was used that $\partial(E/V)/\partial N_\sigma(\mathbf{k})=\partial\Omega/\partial N_\sigma(\mathbf{k})+\mu_\sigma\equiv\epsilon_\sigma(\mathbf{k})$
and for the temperature derivative the Sommerfeld expansion was used,
c.f. Eq.~(\ref{eq:AP_Sommerfeld_Exp}). This result is a trivial generalization of the single species or balanced case, since the effective mass already contains the effect of interactions, such that the contribution to the heat capacity of a single quasiparticle is additive.

\section{Random-phase approximation}\label{sec-RPA}

\begin{figure}
\includegraphics[scale=0.5]{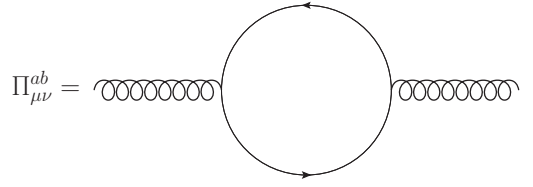}
\caption{The polarization tensor $\Pi_{\mu\nu}^{ab}$ to the first order.}\label{fig-PolarizationTensor}
\end{figure}

\begin{figure}
\includegraphics[scale=0.5]{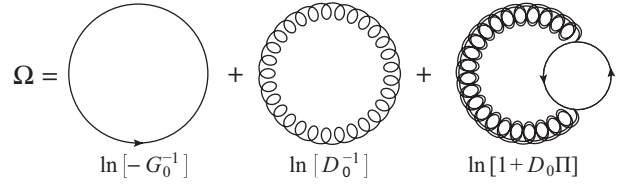}
\caption{The thermodynamic potential in the large-$N_F$ or RPA approximation
where the double gluon line signifies the dressed gluon propagator.}\label{fig-ThermoPotential}
\end{figure}

Our starting point is the partition function
\[Z=\int\mathcal{D}\bar{\psi}\mathcal{D}\psi\mathcal{D}A_\mu\mathcal{D}\bar{\eta}\mathcal{D}\eta\exp\left(-\int\mathrm{d}^4x\mathcal{L}_\mathrm{QCD}^\mathrm{E}\right),\]
containing the Euclidean QCD Lagrangian density fixed in a linear gauge $f_\mu A_\mu^a=0$,
\begin{align*}
\mathcal{L}_{\mathrm{QCD}}^{\mathrm{E}}=&\sum_f\bar{\psi}_f(\gamma_\mu D_\mu+m_f-\gamma_0\mu_f)\psi_f+\frac{1}{4}G_{\mu\nu}^aG_{\mu\nu}^a\\
&+\bar{\eta}^a(\partial_\mu f_\mu\delta^{ab}+gf^{abc}A_\mu^cf_\mu)\eta^b+\frac{1}{2\xi}(f_\mu A_\mu^a)^2,
\end{align*}
where $\psi_f$ is the quark field of flavor $f$ (the color $c$ and spin $s$ degrees of freedom are not shown explicitly) with mass $m_f$ and chemical potential $\mu_f$, $A_\mu^a$ are the gluon fields, $\eta^a$ are the ghost fields, $g$ is the QCD coupling constant, $\xi$ is a gauge fixing parameter, $f^{abc}$ are the fine-structure constants of the color $SU(3)$ group, the covariant derivative $D_\mu=\partial_\mu-igt^aA_\mu^a$ and the antisymmetric gluon field tensor $G^a_{\mu\nu}=\partial_\mu A_\nu^a-\partial_\nu A_\mu^a+gf^{abc}A_\mu^bA_\nu^c$. For large densities, the fermionic degrees of freedom become increasingly important, such that the large-$N_F$ limit could give insight into the behavior of quarks at high densities. Therefore, consider the large-$N_F$ limit with a fixed rescaled coupling $\mathrm{g}^2=g^2N_F$,
\begin{align*}
\mathcal{L}_{\mathrm{QCD}}^{\mathrm{E}}=&\sum_f\bar{\psi}_f\left(\slashed{\partial}+m_f-\gamma_0\mu_f-\mathrm{g}N_F^{-\frac{1}{2}}i\gamma_\mu t^aA_\mu^a\right)\psi_f\\
&+\frac{1}{2}A_\mu^a\left(\partial_\mu\partial_\nu-\partial^2\delta_{\mu\nu}-\frac{1}{\xi}f_\mu f_\nu\right)A_\nu^a\\
&+\bar{\eta}^a\left(\partial_\mu f_\mu\delta^{ab}+\mathrm{g}N_F^{-\frac{1}{2}}f^{abc}A_\mu^cf_\mu\right)\eta^b\\
&+\frac{1}{2}\mathrm{g}N_F^{-\frac{1}{2}}(\partial_{\mu}A_\nu^a-\partial_\nu A_\mu^a)f^{abc}A_\mu^bA_\nu^c\\
&+\frac{1}{4}\mathrm{g}N_F^{-1}f^{abc}f^{ade}A_\mu^bA_\nu^cA_\mu^dA_\nu^e,
\end{align*}
Integrating out the fermions and the ghosts gives
\begin{align*}
\mathcal{L}_{\mathrm{QCD}}^{\mathrm{eff}}=&-\mathrm{Tr}_{c,f,s}\ln\left[-G_{0f}^{-1}\left(1+\mathrm{g}N_F^{-\frac{1}{2}}G_{0f}i\gamma_{\mu}t^aA_\mu^a\right)\right]\\
&-\mathrm{Tr}_c\ln\left(\partial_{\mu}f_\mu\delta^{ab}+\mathrm{g}N_F^{-\frac{1}{2}}f^{abc}A_\mu^cf_\mu\right)\\
&+\frac{1}{2}A_\mu^aD^{-1}_{0,\mu\nu}\delta^{ab}A_\nu^b\\
&+\frac{1}{2}\mathrm{g}N_F^{-\frac{1}{2}}\left(\partial_{\mu}A_\nu^a-\partial_{\nu}A_\mu^a\right)f^{abc}A_\mu^bA_\nu^c\\
&+\frac{1}{4}\mathrm{g}N_F^{-1}f^{abc}f^{ade}A_\mu^bA_\nu^cA_\mu^dA_\nu^e,
\end{align*}
where the subscripts $\{c,f,s\}$ explicitly indicate that the trace should also be taken over color, flavor and spin space. When we expand the first line in $N_F$, the first-order contribution will vanish due to conservation of color, i.e., $\mathrm{Tr}_c[t^a]=0$. Only the second-order expansion in $N_F$ of the first line will give a contribution. Thus expansion in powers of $N_F$ gives
\begin{align*}
\mathcal{L}_{\mathrm{QCD}}^{\mathrm{eff}}=&-\mathrm{Tr}_{c,f,s}\ln(-G_{0f}^{-1})-\mathrm{Tr}_{c}\ln(\partial_\mu f_\mu\delta^{ab})\\
&+\frac{1}{2}A_\mu^a(D_0^{-1}+\sum_f\Pi_f)_{\mu\nu}^{ab}A_\nu^b+\mathcal{O}(N_F^{-\frac{1}{2}}),
\end{align*}
where we defined the polarization tensor $\Pi_{f,\mu\nu}^{ab}$ from the contribution of the quark with flavor $f$ as
\[\Pi_{f,\mu\nu}^{ab}=-g^2\mathrm{Tr}_{c,s}(G_{0f}\gamma_{\mu}t^aG_{0f}\gamma_{\nu}t^b),\]
which is shown diagrammatically in Fig.~\ref{fig-PolarizationTensor}. The polarization tensor $\Pi_{f,\mu\nu}^{ab}$ is diagonal in color space, as can be seen explicitly form Eq.~(\ref{eq:def_Polarization_tensor}). Upon integrating out the gluons the thermodynamic potential density is
\begin{align}
\Omega(T,\{\mu_f\})=&-V^{-1}\beta^{-1}\ln Z\nonumber\\
\simeq&\frac{1}{V\beta}\left\{-N_C\mathrm{Tr}_{f,s}\ln(-G_0^{-1})+\vphantom{\frac{N_G}{2}}\right.\nonumber\\
&+\frac{N_G}{2}\mathrm{Tr}\left[\ln(D_0^{-1})-2\ln(\partial_{\mu}f_{\mu})-\ln\frac{1}{\xi}\right]\nonumber\\
&\left.+\frac{N_G}{2}\mathrm{Tr}\ln\left(1+D_0\sum_f\Pi_f\right)\right\},\label{eq:Therm_pot}
\end{align}
where $N_C$ is the number of colors, $N_G=N_C^2-1$ is the number of gluons, and we use the QCD values $N_C=3$ and $N_G=8$ in the following. The first two terms are the ideal Fermi and Bose gas contributions to the thermodynamic potential density, where the term $\mathrm{Tr}\ln(\partial_{\mu}f_\mu)$
cancels the two unphysical degrees of freedom from $\mathrm{Tr}\ln(D_0^{-1})$.
The last term is the RPA correction (ring sum) to the thermodynamic
potential density, as is shown in Fig~\ref{fig-ThermoPotential}. Since we are mainly interested in the temperature dependence of $\Omega$ and the fact that the above definition contains (divergent) zero-temperature contributions, the $T=0$ expression will be subtracted, i.e., we consider $\Delta\Omega(T,\{\mu_f\})\equiv\Omega(T,\{\mu_f\})-\Omega(0,\{\mu_f\})$.
In Sec.~\ref{sec-weakcoupling} an example is considered to check the expression for the Landau effective mass analytically and in Sec.~\ref{sec-thermpotential} the above thermodynamic potential density is calculated numerically for the case of a two-flavor imbalanced quark system.

\section{Fermi-liquid parameters in weak coupling}\label{sec-weakcoupling}

In the limit of weak coupling the interaction contribution to the
thermodynamic potential density is given solely by the exchange diagram
\[\Omega_{\textrm{int}}=\frac{N_G}{2V\beta}\sum_f\mathrm{Tr}(D_0\Pi_f).\]
Using the definition of the free gluon propagator in the Lorentz gauge, this can be written as
\begin{align*}
\Omega_{\textrm{int}}&=\frac{N_G}{2\beta}\sum_f\sum_{\omega_q}\int\frac{\mathrm{d}^3q}{(2\pi)^3}\frac{\Pi_{f,\mu\mu}(i\omega_q,\mathbf{q})}{\omega_q^2+q^2}\\
&=g^2\frac{N_G}{4}\sum_f\sum_{s_1,s_2,s_3=\pm1}\int\frac{\mathrm{d}^3p\mathrm{d}^3q}{(2\pi)^6}\mathcal{F}^f_{s_1,s_2,s_3}(\mathbf{p},\mathbf{q})\\
&\quad\times[N_f^{s_2}(\mathbf{q})N_B^{s_3}(\mathbf{p}-\mathbf{q})(1-N_f^{s_1}(\mathbf{p}))\\
&\qquad-N_f^{s_1}(\mathbf{p})(1-N_f^{s_2}(\mathbf{q}))(1+N_B^{s_3}(\mathbf{p}-\mathbf{q}))],
\end{align*}
where the Fermi and Bose distributions are defined as
\[N_f^s(\mathbf{p})=\frac{1}{e^{\beta[s\epsilon_{0f}(\mathbf{p})-\mu_f]}+1},\quad N_B^s(\mathbf{p})=\frac{1}{e^{\beta[s\epsilon^g(\mathbf{p})]}-1},\]
respectively, the explicit expression for $\Pi_{f,\mu\mu}$ is shown in Eq.~(\ref{PolarizationTensor}), the Matsubara sum over the bosonic frequencies $\omega_q$ was performed, $\epsilon_{0f}(\mathbf{p})=\sqrt{p^2+m_f^2}$ denotes the free quark dispersion while $\epsilon^g(\mathbf{p})=p$ is the free gluon dispersion, and the function $\mathcal{F}$ was defined as
\begin{align}
\mathcal{F}^f_{s_1,s_2,s_3}(\mathbf{p},\mathbf{q})=&\frac{s_3}{\epsilon_{0f}(\mathbf{p})\epsilon_{0f}(\mathbf{q})\epsilon^g(\mathbf{p}-\mathbf{q})}\nonumber\\
&\times\frac{-\epsilon_{0f}(\mathbf{p})\epsilon_{0f}(\mathbf{q})+s_1s_2(\mathbf{p}\cdot\mathbf{q}+2m^2)}{s_1\epsilon_{0f}(\mathbf{p})-s_2\epsilon_{0f}(\mathbf{q})-s_3\epsilon^g(\mathbf{p}-\mathbf{q})},\label{functionf}
\end{align}
which will soon be shown to play the role of the interaction between
two quarks ($s_1=s_2=1$), two anti-quarks ($s_1=s_2=-1$)
or a quark and an antiquark ($s_1=-s_2=1$ or $s_1=-s_2=-1$)
mediated by the emission or absorption of a gluon (depending on the
sign of $s_3$ relative to $s_1$ and $s_2$).

The quasiparticle energy and effective interaction can be obtained by varying the thermodynamic potential with respect to the Fermi distribution. In the flavor-imbalanced case, the distributions depend only on the flavor but not on the color and spin indices. To distinguish between the different contributions for each particle species it can be checked that the multiplication factor $N_G/4$ could be written as $\sum_{c,c',s,s'}t_{cc'}^at_{c'c}^a/8$, where $s$ and $s'$ are just dummy indices denoting the spin degrees of freedom, which are helpful to arrive at the following expressions. Note that $\Omega_\mathrm{int}$ is independent of the spin index, and in fact the spin degrees of freedom has already been summed over in the expression for $\mathcal{F}$. Explicitly we find for these quantities
\begin{widetext}
\begin{align}
&\epsilon_{c,f,s}(\mathbf{k})-\mu_f\equiv\frac{\delta\Omega}{\delta N_{c,f,s}^+(\mathbf{k})}=[\epsilon_{0f}(\mathbf{k})-\mu_f]-\sum_{s'}\frac{g^2}{4}\sum_{c'}t_{cc'}^at_{c'c}^a\sum_{s_2,s_3=\pm1}\int\frac{\mathrm{d}^3p}{(2\pi)^3}\mathcal{F}^f_{+,s_2,s_3}(\mathbf{k},\mathbf{q})\nonumber\\
&\qquad\qquad\qquad\qquad\qquad\qquad\qquad\qquad\qquad\qquad\qquad\times[(1+N_B^{s_3}(\mathbf{k}-\mathbf{q})(1-N_f^{s_2}(\mathbf{q}))+N_B^{s_3}(\mathbf{k}-\mathbf{q}))N_f^{s_2}(\mathbf{q})],\label{Quasiparticle-energy}\\
&f_{\{c,f,s\};\{c',f',s'\}}(\mathbf{k},\mathbf{k}')\equiv\frac{\delta^2\Omega}{\delta N_{c,f,s}^+(\mathbf{k})\delta N_{c',f',s'}^+(\mathbf{k}')}=\frac{g^2}{4}t_{cc'}^at_{c'c}^a\delta_{ff'}\sum_{s_3=\pm1}\mathcal{F}^f_{+,+,s_{3}}(\mathbf{k},\mathbf{k}'),\label{Quasiparticle-int}
\end{align}
\end{widetext}
where $\sum_{c'}t_{cc'}^at_{c'c}^a=\frac{N_G}{2N_C}$ and $t_{cc'}^at_{c'c}^a=1/2-\delta_{cc'}/2N_C$, according to the definitions in Appendix~\ref{app-Euclidean}.

In the quasiparticle energy Eq.~(\ref{Quasiparticle-energy}) the first term is the noninteracting part, while the second term corresponds to the quark self-energy, as is shown in Appendix~\ref{app-quark-selfenergy}. The angular averaged interaction parameters are defined by the coefficients of the Legendre polynomial expansions of the effective interaction Eq.~(\ref{Quasiparticle-int}) evaluated on the Fermi sphere, by denoting $\cos\theta=\hat{\mathbf{k}}\cdot\hat{\mathbf{k}}'$,
\begin{widetext}
\begin{align*}
f_{l;\{c,f,s\}}&=\sum_{c',f',s'}f_{l;\{c,f,s\};\{c',f',s'\}}=\frac{2l+1}{2}\sum_{c',f',s'}\int\mathrm{d}\theta\sin\theta f_{\{c,f,s\};\{c',f',s'\}}(\hat{\mathbf{k}}k_f,\hat{\mathbf{k}}'k_{f'})P_l(\cos\theta),
\end{align*}
\end{widetext}
which gives for the zeroth and first interaction parameters
\begin{align}
f_{0;f}&=\frac{g^2N_GN_S}{8N_C}\frac{1}{\mu_{0f}^2}\left[1+\frac{\lambda^2+2m_f^2}{4k_f^2}\ln\left(\frac{\lambda^2}{\lambda^2+4k_f^2}\right)\right],\nonumber\\
f_{1;f}&=\frac{g^2N_GN_S}{8N_C}\frac{3}{\mu_{0f}^2}\frac{\lambda^2+2m_f^2}{2k_f^2}\nonumber\\
&\qquad\qquad\times\left[1+\frac{\lambda^2+2k_f^2}{4k_f^2}\ln\left(\frac{\lambda^2}{\lambda^2+4k_f^2}\right)\right],\label{WeakcoupLandauparam}
\end{align}
where $N_S=2$ is the number of spin degrees of freedom, and a regulatory gluon mass $\lambda$ was introduced, i.e., $\epsilon_k^g=\sqrt{k^2+\lambda^2}$. We see that in this approximation the interaction is color- and spin-independent. For the sake of simplicity, therefore, we from now on omit the unnecessary color and spin indices, and use $k_f$ and $\mu_{0f}$ for the Fermi momentum and the non-interacting Fermi energy of the quark with flavor $f$. The effective mass of the balanced case, Eq.~(\ref{rel-Landaumass}), can be obtained from the interaction parameters Eq.~(\ref{WeakcoupLandauparam}), and Eq.~(\ref{LandauFermiparameter}), by expanding to the first order in the coupling constant
\begin{equation}\label{Effectivemassexpand}
m^*=\mu+\sum_{\sigma'}\frac{\mu}{3}\frac{k_Fm^*}{2\pi^2}f_{1,\sigma\sigma'}=\mu+\frac{k_F\mu_0^2}{6\pi^2}f_1+\mathcal{O}(g^4),
\end{equation}
where it was used that $\mu=\mu_0+\mathcal{O}(g^2)$ as follows from
Eq.~(\ref{Quasiparticle-energy}). For the imbalanced case, a similar result can be obtained for each species according to Eq.~(\ref{imbalancemass}),
\begin{equation}\label{imbalancemassexpand}
m_f^*=\mu_f+\frac{\mu_ff_{1;f}(k_f)N_f(0)}{3}\approx\mu_f+\frac{k_f\mu_{0f}^2}{6\pi^2}f_{1;f}(k_f),
\end{equation}
where the Kronecker delta $\delta_{ff'}$ in the interaction simplifies the expression considerably, such that we obtain the similar expression as in Eq.~(\ref{Effectivemassexpand}).

As a special case, we can use the above results to discuss a two-flavor imbalanced system, where the average effective mass and chemical potential can be defined as $m^*=(m_+^*+m_-^*)/2$ and $\mu=(\mu_++\mu_-)/2$ with the subscripts $+$ and $-$ for the majority and minority flavors, respectively. The imbalance can be quantified as $h=(\mu_+-\mu_-)/(\mu_++\mu_-)$. For small imbalance, the factors $k_\pm f_{1;\pm}$ can be expanded in $h$ with respect to the value in the balanced case, namely $k_\pm f_{1;\pm}(k_\pm)\approx k_Ff_1(k_F)(1\pm\Delta_{1\pm}h+\Delta_{2\pm}h^2)$. Note that, the linear coefficients of the expansion can, in general, be different for each flavor. We obtain the expression of $m^*$ for small $h$
\begin{align*}
m^*=&\mu+\frac{k_F\mu^2f_1(k_F)}{12\pi^2}[2+(\Delta_{1+}-\Delta_{1-})h\\ &+(2+2\Delta_{1+}+2\Delta_{1-}+\Delta_{2+}+\Delta_{2-})h^2+\mathcal{O}(h^3)],
\end{align*}
where the $h^2$ term is kept since the linear term may vanish for the symmetric case, namely $\Delta_{1+}=\Delta_{1-}$, and we have replaced $\mu_0$ with $\mu$ because the difference is in higher order of $g$. Compared with Eq.~(\ref{averageimbalancemass}), the effective Landau parameter for the average effective mass is
\begin{align*}
F_1=&\frac{k_F\mu f_1(k_F)}{4\pi^2}[2+(\Delta_{1+}-\Delta_{1-})h\\ &+(2+2\Delta_{1+}+2\Delta_{1-}+\Delta_{2+}+\Delta_{2-})h^2+\mathcal{O}(h^3)].
\end{align*}
Similarly, we can obtain the effective mass difference $\Delta_{m^*}=(m_+^*-m_-^*)/2$ as a function of $h$
\begin{align*}
\Delta_{m^*}=&\mu h+\frac{k_F\mu^2f_1(k_F)}{12\pi^2}[(4+\Delta_{1+}+\Delta_{1-})h\\ &+(2\Delta_{1+}-2\Delta_{1-}+\Delta_{2+}-\Delta_{2-})h^2+\mathcal{O}(h^3)].
\end{align*}
It is quite natural to find $\Delta_{m^*}\propto h$ in the leading order.

The consistency of the above results will now be verified to the lowest order in the coupling constant using the quark self-energy. Starting from the dispersion and self-energy of Eq.~(\ref{Quasiparticle-energy}),
the effective mass in Eq.~(\ref{Effectivemassexpand}) or Eq.~(\ref{imbalancemassexpand}) can also be derived in a different manner. The quasiparticle pole of the quark propagator is
\[\epsilon_f(\mathbf{p})\equiv\epsilon_{0f}(\mathbf{p})+\Sigma_{c,f,s}^+(\epsilon_{0f}(\mathbf{p})-\mu,\mathbf{p}),\]
where $\Sigma^+$ is the renormalized positive-energy projected self-energy. The effective mass corresponding to this pole is most easily defined from the Fermi velocity of the quasiparticle
\[\frac{m_f^*}{\mu_f}\equiv\frac{k_f}{v_f^*\epsilon_f}=k_f\left[\frac{1}{\epsilon_f(\mathbf{p})}\left(\frac{\partial\epsilon_f(\mathbf{p})}{\partial p}\right)^{-1}\right]_{p=k_f}.\]
Inserting the definition of the quasiparticle pole into the above gives, using $\partial\epsilon_{0f}(\mathbf{p})/\partial p=p/\epsilon_{0f}(\mathbf{p})$,
\begin{align}
m_f^*=&\mu_fk_f\left[\epsilon_{0f}(\mathbf{p})\frac{\partial\epsilon_{0f}(\mathbf{p})}{\partial p}\right.\nonumber\\ &\qquad\left.+\frac{\partial\epsilon_{0f}(\mathbf{p})\Sigma_{c,f,s}^+(\epsilon_{0f}(\mathbf{p})-\mu_f,\mathbf{p})}{\partial p}\right]_{p=k_f}^{-1}+\mathcal{O}(g^4)\nonumber\\
=&\mu_f-\left[\frac{\partial\epsilon_{0f}(\mathbf{p})\Sigma_{c,f,s}^+(\epsilon_{0f}(\mathbf{p})-\mu_f,\mathbf{p})}{\partial\epsilon_{0f}(\mathbf{p})}\right]_{p=k_f}+\mathcal{O}(g^4).\label{effectivemassnew}
\end{align}
Using that the explicit form of the self-energy in the zero-temperature
limit can be written as (c.f. Eq.~(\ref{quarkselfenergy}))
\begin{align*}
\Sigma_{c,f,s}^+(\epsilon_{0f}(\mathbf{p})-\mu_f,\mathbf{p})=&\\ \sum_{c',f',s'}\int\frac{\mathrm{d}^3q}{(2\pi)^3}&f_{\{c,f,s\};\{c',f',s'\}}(\mathbf{p},\mathbf{q})N_{f'}^+(\mathbf{q}),
\end{align*}
and the relation Eq.~(\ref{eq:AP_eff_int_transf}) between the vector
derivatives of the effective interaction, the derivative in Eq.~(\ref{effectivemassnew}) can be rewritten as
\begin{widetext}
\begin{align*}
&\left[\frac{\partial p}{\partial\epsilon_{0f}(\mathbf{p})}\frac{\partial\epsilon_{0f}(\mathbf{p})\Sigma_{c,f,s}^+(\epsilon_{0f}(\mathbf{p})-\mu_f,\mathbf{p})}{\partial p}\right]_{p=k_f}=\frac{\mu_{0f}}{k_f}\left[\sum_{c',f',s'}\int\frac{\mathrm{d}^3q}{(2\pi)^3}\hat{\mathbf{p}}\cdot\frac{\partial\epsilon_{0f}(\mathbf{p})f_{\{c,f,s\};\{c',f',s'\}}(\mathbf{p},\mathbf{q})}{\partial\mathbf{p}}N_{f'}^+(\mathbf{q})\right]_{p=k_f}\\
=&-\frac{\mu_{0f}}{k_f}\int\frac{\mathrm{d}q}{2\pi^2}q^2\epsilon_{0f}(\mathbf{q})\delta(q-k_{f})\frac{1}{3}\left[3\int\frac{\mathrm{d}\hat{\mathbf{q}}}{4\pi}\hat{\mathbf{p}}\cdot\hat{\mathbf{q}}\sum_{c',f',s'}f_{\{c,f,s\};\{c',f',s'\}}(\mathbf{p},\mathbf{q})\right]_{p=k_f} =-\frac{k_f\mu_{0f}^2}{2\pi^2}\frac{f_{1;\{c,f,s\}}(k_f)}{3},
\end{align*}
\end{widetext}
where partial integration was used and $\delta_{ff'}\partial_{\mathbf{q}}N_{f'}^+(\mathbf{q})=-\delta_{ff'}\hat{\mathbf{q}}\delta(q-k_f)$
at zero temperature. Therefore Eq.~(\ref{effectivemassnew}) is identical to Eq.~(\ref{imbalancemassexpand}), which shows that the phenominological Landau argument is indeed consistent with the microscopic diagrammatic calculation.

\section{Dressed gluon propagator}\label{sec-gluonpropagator}

Before we start the calculation of the temperature dependence of the large-$N_F$ thermodynamic potential, it is useful to examine the dressed gluon propagator, which is a crucial ingredient of the RPA theory. The most physical gauge to study the propagator is the Coulomb gauge, since its form results from considering linear response \cite{Kapusta:2006,Le Bellac:1996}. This is also by far the most used gauge in condensed-matter theory. In this gauge the gluon propagator is \cite{Le Bellac:1996}
\begin{align*}
D_{\mu\nu}(Q)=&\frac{P_{\mu\nu}^T}{Q^2+G(Q)}+\frac{Q^2}{q^2}\frac{\delta_{\mu0}\delta_{\nu0}}{Q^2+F(Q)}+\frac{\xi Q^2}{q^4}\frac{Q_\mu Q_\nu}{Q^2}\\
\equiv&D^T(Q)P_{\mu\nu}^T-D^L(Q)\delta_{\mu0}\delta_{\nu0}+\frac{\xi Q^2}{q^4}\frac{Q_\mu Q_\nu}{Q^2},
\end{align*}
where $P_{\mu\nu}^T$ is the three-dimensional transverse projector defined in Appendix~\ref{app-Green}, $F$ and $G$ are related to the longitudinal and transverse projections of $\Pi_{\mu\nu}$ including the contributions of various species as well as the vacuum. More details can be found in Appendix~\ref{app-gluon-selfenergy}.

\begin{figure}
\resizebox{1\linewidth}{!}{\includegraphics{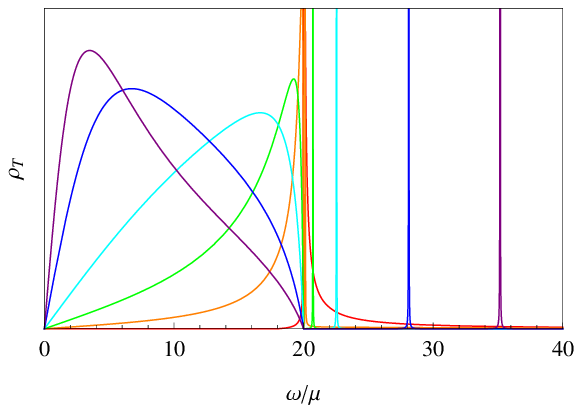}}
\resizebox{1\linewidth}{!}{\includegraphics{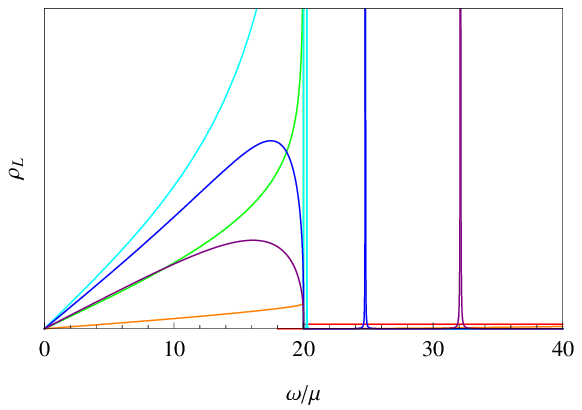}}
\caption{(color online). The spectral function of the transverse (upper panel) and longitudinal (lower panel) gluon propagators as a function of frequency $\omega$ for several values of $\mu$ in the balanced case. Shown here for $T=0$, $m=1$, $q=20$, $g=1/2$, $N_F=2$ and for different curves $\mu=1, 20, 50, 100, 200, 300$ from red to purple.\label{fig:rhoTL}}
\end{figure}

\begin{figure}
\resizebox{1\linewidth}{!}{\includegraphics{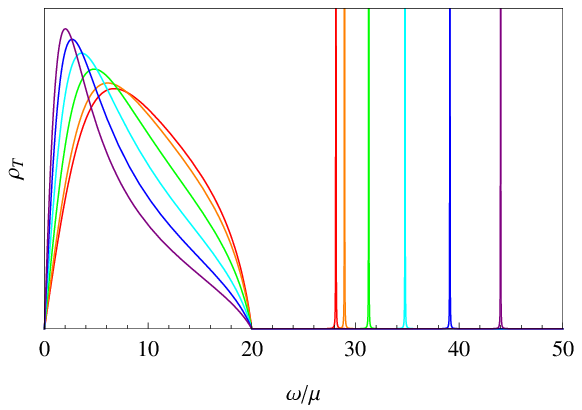}}
\resizebox{1\linewidth}{!}{\includegraphics{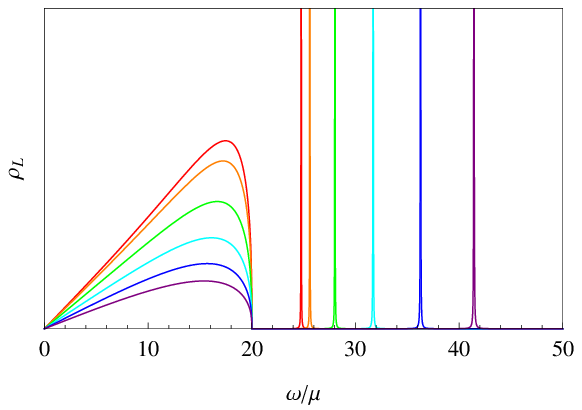}}
\caption{(color online). The spectral functions of the transverse (upper panel) and longitudinal (lower panel) gluons as functions of $\omega$ for various $T$ in the balanced case, with $\mu=200$, $m=1$, $q=20$, $g=1/2$, $N_F=2$ and for different curves $T=1, 40, 80, 120, 160, 200$ from red to purple. (In principle, $T$ should be much smaller than $\mu$, but here we show also large values of $T$ to demonstrate the curves more clearly.)\label{fig:rhoTLtemp}}
\end{figure}

Consider the spectral functions of the transverse and longitudinal
propagator, which are defined by
\[\rho^{T,L}(\omega,\mathbf{q})\equiv\frac{1}{\pi}\Im\left[D^{T,L}(\omega+i0,\mathbf{q})\right].\]
In general the spectral function depends on the gauge, however, the
positions of the poles are gauge independent, and the Coulomb gauge
has the additional property $\rho^{T,L}(\omega,\mathbf{q})>0$ for $\omega>0$
as required of a physical spectral function. For the balanced case, the spectral functions are shown in Fig.~\ref{fig:rhoTL} for several values
of $\mu$ and in Fig.~\ref{fig:rhoTLtemp} for several
values of $T$. The latter includes a small-$T$ correction to $F$
and $G$, see Eq.~(\ref{eq:AP_Pi_temperature_correction})
\footnote{Due to the inclusion of a small $T$-correction the spectral functions
will no longer satisfy $\rho^{T,L}>0$.}. The transverse and longitudinal plasmon modes and the large contribution due to the decay of the gluon into the particle-hole
continuum ($0<\omega<q$) are clearly visible. Note that massless quarks are used in the limit of high density, c.f. Appendix~\ref{app-gluon-selfenergy}. However, a small but finite quark mass is necessary to renormalize the real part of the polarization tensor of the vacuum, as shown in Eq.~(\ref{eq:AP_renormalised_vacuum}). Since the vacuum contribution plays no significant role in the following calculation, we simply take this nonzero quark mass $m$ in the vacuum term the same for different flavors.

\begin{figure}
\resizebox{1\linewidth}{!}{\includegraphics{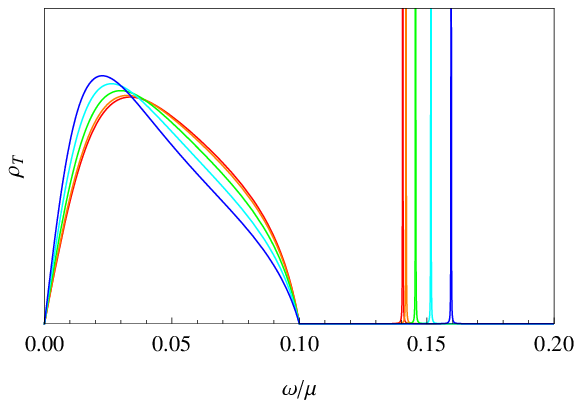}}
\resizebox{1\linewidth}{!}{\includegraphics{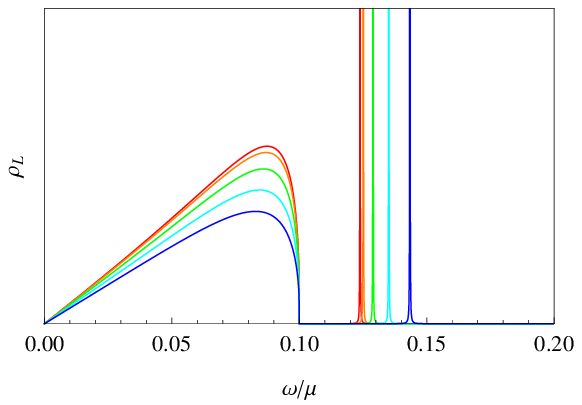}}
\caption{(color online). The spectral functions of the transverse (upper panel) and longitudinal (lower panel) gluons as functions of $\omega$ for various $h$, with $T=0$, $m=0.001$, $q=0.1$, $\mu=1$, $g=1/2$ and for different curves $h=0, 0.2, 0.4, 0.6, 0.8$ from red to blue.\label{fig:rhoTLh}}
\end{figure}

For the two-flavor imbalanced case, it is clear from Eq.~(\ref{eq:Therm_pot}) that in the leading-order correction each flavor contributes separately to the polarization tensor. Because of the above setting, there is no mass imbalance in the present system, therefore it is enough to consider only positive $h$ due to the symmetry. Following the notation in Sec.~\ref{sec-weakcoupling}, all the above results can be easily generalized to such an imbalanced system by using $\mu(1+h)$ to replace $\mu_+$, and $\mu(1-h)$ for $\mu_-$. As $h\rightarrow0$, the system reduces to the balanced case. Now the low temperature condition requires $T\ll\mu(1\pm h)$, such that $h$ can not be too close to $1$, namely the extremely imbalanced case. The spectral function for a two-flavor imbalanced system is shown for various $h$ in Fig.~\ref{fig:rhoTLh}. Comparing with Fig.~\ref{fig:rhoTLtemp}, we see that increasing $h$ has a similar effect as increasing $T$.

\begin{figure}
\resizebox{1\linewidth}{!}{\includegraphics{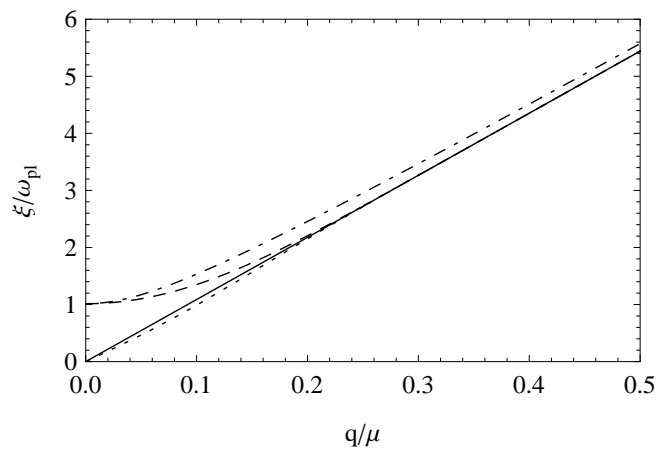}}
\caption{The solutions of Eq.~(\ref{eq:Re_InvProp_zero}) at $T=h=0$, with the dot-dashed curve for $\xi^T$, the dashed curve for $\xi_2^L$, and dotted curve for $\xi_1^L$. The solid line is a guide to the eye and corresponds to the dispersion of undressed gluons, $\omega=q$.\label{fig:Dispersion}}
\end{figure}

The dispersion relations of the modes can be found by solving
\begin{equation}
\Re\left[D_{T,L}^{-1}(\omega,q)\right]=0.\label{eq:Re_InvProp_zero}
\end{equation}
In the transverse case there is a single solution $\xi^T(q)>q$, however, in the longitudinal case there are two solutions $\xi_1^L(q)<q<\xi_2^L(q)$, as shown in Fig.~\ref{fig:Dispersion}. Note that $\xi_1^L(q)$ is not a real propagating mode because the imaginary part in the region $0<\omega<q$ is large due to particle-hole creation processes, as explained in Appendix~\ref{app-gluon-selfenergy}. The two plasmon modes $\xi^T(q)$ and $\xi^L_2(q)$ approach the so-called plasma frequency $\omega_\textrm{pl}$ as $q\rightarrow0$. The plasma frequency of the transverse and longitudinal mode can be found by expanding the inverse propagators for small $q$ and small $\omega$, which in the zero-temperature limit gives for both cases
\[1-\frac{2}{3}\frac{m_g^2}{\omega^2}\left(1-\frac{\omega^2}{4\mu^2}\ln\frac{\omega^2}{4\mu^2}\right)=0,\]
whose solution for small coupling constant yields $\omega_\textrm{pl}\simeq\sqrt{2/3}m_g$ \cite{Le Bellac:1996}. Here $m_g$ is the gluon thermal mass whose expression is obtained in the hard dense and hard thermal loop approximation as shown in Eq.~(\ref{thermalmass}). In the limit $q\gg m_g\sim g\mu$ all solutions reduce to $\omega=q$, see Eq.~(\ref{eq:AP_FG_large_pomega_behavior}).

Furthermore, for the limit $\omega,q\gg\mu$ the vacuum becomes increasingly dominant
such that Eq.~(\ref{eq:Re_InvProp_zero}) has a zero at large $\mathrm{Q}$
called the Landau pole, $\mathrm{Q}^2=\exp\left(\frac{5}{3}+\frac{24\pi^2}{g^2N_F}\right)m^2\equiv\Lambda_L^2$
\cite{Ryder:1996}, c.f. Eq.~(\ref{eq:AP_FG_large_pomega_behavior}), which, however, plays no role for our purpose as we are interested in the low-temperature behavior of the theory that is hardly influenced by the high-energy behavior of the gluon propagator. The dispersion relations for the case with nonzero $T$ or $h$ is similar but just with a little higher $m_{\textrm{pl}}$. In fact, as a generalization of Eq.~(\ref{thermalmass}), the thermal mass of a two-flavor imbalanced quark system reads
\begin{equation*}
m_g^2=\frac{g^2}{4\pi^2}\left[\mu^2(1+h)^2+\mu^2(1-h)^2+\frac{2\pi^2T^2}{3}\right],
\end{equation*}
where $h$ thus plays the same role as $\pi T/\sqrt{3}\mu$.

\section{The full large-$N_F$ thermodynamic potential}\label{sec-thermpotential}

The full ideal-gas contribution to the thermodynamic potential density with the zero-temperature contribution subtracted is (c.f. Appendix~\ref{app-thermpotential})
\[\Delta\Omega_0=-N_C\left(N_F\frac{7\pi^2T^4}{180}+\sum_f\frac{T^2\mu_f^2}{6}\right)-N_G\frac{T^4}{45\pi^2}.\]
The first part is the contribution of an ideal massless Fermi gas, while the second part is the Stefan-Boltzmann law of an ideal Bose gas. The RPA correction to $\Omega$, the last term in Eq.~(\ref{eq:Therm_pot}), can be written as
\begin{align*}
\Omega_{\mathrm{RPA}}(T,\{\mu_f\})&=\frac{N_G}{2V\beta}\sum_{\omega_n,q}\mathrm{Tr}\ln(1+D_0\Pi)\\ &=\frac{N_G}{2V\beta}\sum_{\omega_n,q}\ln\mathrm{Det}\left(1+\frac{FP^L}{Q^2}+\frac{GP^T}{Q^2}\right)\\ &=\frac{N_G}{2V\beta}\sum_{\omega_n,q}\ln\left(1+\frac{F}{Q^2}\right)\left(1+\frac{G}{Q^2}\right)^2,
\end{align*}
where the Lorentz gauge is used, while for the Coulomb gauge the second identity is not valid but the last result is still the same, which is a consequence of the gauge invariance of the thermodynamic potential density. Using contour deformations to carry out the Matsubara sum, as shown in Appendix~\ref{app-Matsubara}, we obtain
\begin{align*}
\Omega_{\mathrm{RPA}}(T,\{\mu_f\})&=\frac{N_G}{2\pi}\int\frac{\mathrm{d}q\mathrm{d}\omega}{(2\pi)^3}4\pi q^2[2N_B(\omega)+1]\\ &\times\{\Im[\ln\tilde{F}(\omega_+,\mathbf{q})]+2\Im[\ln\tilde{G}(\omega_+,\mathbf{q})]\},
\end{align*}
where $\tilde{F}(\omega_+,\mathbf{q})=1+F(\omega_+,\mathbf{q})/(-\omega_+^2+q^2)$, and similar for $\tilde{G}$ by replacing $F$ with $G$. The temperature dependence can be obtained by subtracting the zero-temperature contribution. Since $F$ and $G$ contain corrections of order $T^2$, the leading-order correction in $\Delta\Omega_{\mathrm{RPA}}$ comes from two parts,
\begin{align*}
\Delta\Omega&_{\mathrm{RPA}}(T,\{\mu_f\})=\frac{N_G}{2\pi^3}\int\mathrm{d}q\mathrm{d}\omega q^2\left[\vphantom{\frac{1}{2}}N_B(\omega)\right.\\ &\left(\arctan\frac{\Im(\tilde{F}^0)}{\Re(\tilde{F}^0)}+\pi\Theta[-\Re(\tilde{F}^0)]\mathrm{sgn}[\Im(\tilde{F})^0]\right.\\ &\left.+2\arctan\frac{\Im(\tilde{G}^0)}{\Re(\tilde{G}^0)}+2\pi\Theta[-\Re(\tilde{G}^0)]\mathrm{sgn}[\Im(\tilde{G}^0)]\right)\\
&\left.\frac{1}{2}[\Im(\ln\tilde{F}^T)-\Im(\ln\tilde{F}^0)+2\Im(\ln\tilde{G}^T)-2\Im(\ln\tilde{G}^0)]\vphantom{\frac{1}{2}}\right], \end{align*}
where the superscript $T$ or $0$ means the corresponding terms are taken at nonzero $T$ or $T=0$. We will refer to the first part as the $N_B$ term and the second as the non-$N_B$ term. The leading correction from the non-$N_B$ term can be shown to be proportional to $T^2$ and is not of great interest in our study, since we will concentrate on the anomalous and dominant $T$ dependence, which is a consequence of the $N_B$ term. In the $N_B$ term, the arctangent terms can be interpreted as contributions due to production and decay of thermal gluons, because of their dependence on the imaginary part of the gluon self-energies $F$ and $G$, while the theta function terms are interpreted as a correction to the ideal gas law due to thermal plasmon modes.

The frequency integral over the theta function can be performed explicitly.
In both the transverse and longitudinal case the sign of the imaginary
part is positive when the real part is negative, such that after the
frequency integration the result of the integral is proportional to
\[T^4\int_0^\infty x^2\ln\frac{1-e^{-\beta\xi_2(\beta^{-1}x)}}{1-e^{-\beta\xi_1(\beta^{-1}x)}}\mathrm{d}x,\]
where $x=q/T$ and $\xi_1<\omega<\xi_2$ signifies the region where $\Re_{F,G}<0$. In the limit of low temperature this integral will go to a constant. The theta-function contribution can thus be neglected since it is of higher order in the temperature than is of interest to us here. 

\begin{figure}
\resizebox{1\linewidth}{!}{\includegraphics{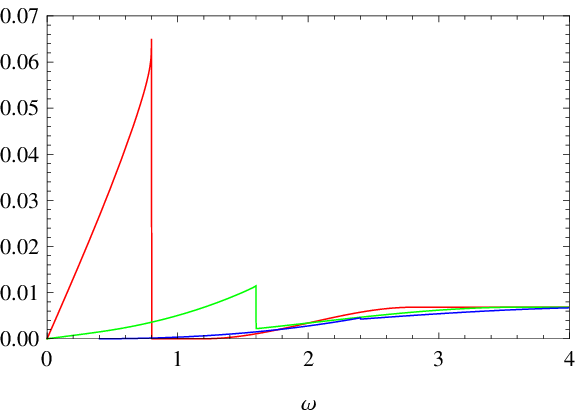}}
\resizebox{1\linewidth}{!}{\includegraphics{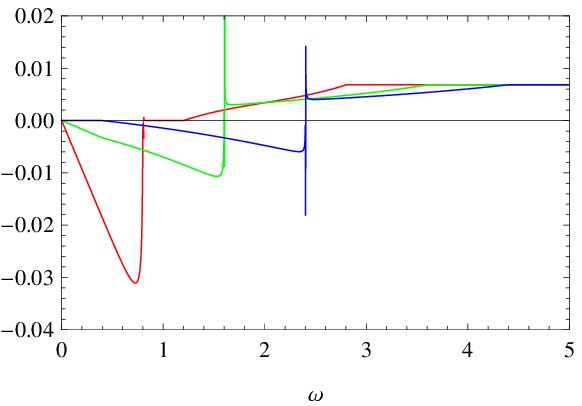}}
\caption{(color online). The structure of $\arctan(\Im_F/\Re_F)$ (upper panel) and $\arctan(\Im_G/\Re_G)$ (lower panel) with $T=0$, $m=0.001\mu$, $h=0$, $g=1/2$, and for different curves, $q/\mu=0.8$ (red), $1.6$ (green), and $2.4$ (blue), respectively. The contribution for $\max(0,q-2\mu)<\omega<q$ is solely due to particle-hole contributions, while the contribution $q<\omega<\infty$ is due to particle-antiparticle processes (a combination of finite density and vacuum processes). The discontinuities at $\omega=q$ correspond to the plasmon modes shown in Fig.~\ref{fig:Dispersion}. The constant tails are due to the vacuum contribution, which, however, will not cause divergence because of the Bose distribution function $N_B(\omega)$.\label{fig:ArcTan}}
\end{figure}

Next we perform the integrals over the arctangents, whose structure at $T=h=0$ is shown in Fig.~\ref{fig:ArcTan}. Note that, for the study of the leading-order $T$ corrections, it is not necessary to include the $T^2$ term in $F$ and $G$ since the integral with $N_B(\omega)$ at low $T$ is already in the order of $T^2$.
The dominant contribution for small temperatures ($T\ll\mu$) comes from the frequency integration over the domain $\omega\in[0,q]$ for the case $q<2\mu$, which is due to particle-hole creation. For the two-flavor balanced case ($N_F=2$, $h=0$), it was found numerically that the integral in the limit of small temperatures is
\[\frac{g^2\mu^2T^2}{\pi^2}(c_1^L-c_2^L\ln g^2)\]
for the electric (longitudinal) gluons and
\[2\frac{g^2\mu^2T^2}{\pi^2}\left(-c_1^T+c_2^T\ln\frac{g^2T}{\mu}\right)\]
for the magnetic (transverse) gluons, with $c_1^L\simeq0.48$,
$c_2^L\simeq0.16$, $c_1^T\simeq0.42$, and $c_2^T\simeq0.056$. These results are quite close to the analytic results for the leading-order correction terms at small $g$ and $T$ obtained in Ref.~\cite{Ipp:2004b}, where $c_1^L=[\ln(4\pi^2)-1]/6\approx0.4460$, $c_2^L=1/6\approx0.1667$, $c_1^T=[\gamma_E-6\zeta'(2)/\pi^2+3/2+\ln(32\pi)]/18\approx0.4032$, and $c_2^T=1/18\approx0.05556$.

For the imbalanced case, the above result for the transverse gluons can be generalized to
\begin{equation}\label{CT}
\frac{1}{2}\sum_{s=\pm1}2\frac{g^2\mu^2(1+sh)^2T^2}{\pi^2}\left[-c_1^T+c_2^T\ln\frac{g^2T}{\mu(1+sh)}\right],
\end{equation}
where the contributions from both flavors with chemical potentials $\mu(1\pm h)$ are additive. The longitudinal part needs some further discussion. Since the static long-wavelength longitudinal modes are screened (c.f. Appendix~\ref{app-gluon-selfenergy}), the two chemical potentials contributing to $\Re_F$ can not be separated even in the lowest-order term. This is different from the result obtained in weak coupling, as shown in Eq.~(\ref{Quasiparticle-int}), where no interaction between the two flavors is involved. Therefore it is not surprising to find that a simple generalization of the balanced case, as $\sum_{s=\pm1}g^2\mu^2(1+sh)^2T^2(c_1^L-c_2^L\ln g^2)/2\pi^2$, does not fit well with the numerical results. To obtain a reasonable ansatz for the longitudinal part, we integrate the corresponding arctangent term of the imbalanced case in the low temperature limit up to $\mathcal{O}(g^2)$ to obtain
\begin{equation}\label{CL}
\frac{1}{2}\sum_{s=\pm1}\frac{g^2\mu^2(1+sh)^2T^2}{\pi^2}\left[\frac{1}{6}\ln\frac{4\pi^2(1+sh)^2}{g^2(1+h^2)}-\frac{1}{6}\right],
\end{equation}
where we see that the factor $(1+h^2)=\frac{1}{2}\sum_{s=\pm1}(1+sh)^2$ is a mixture effect of the two flavors. This expression fits the numerical results very well with an error of only $3.5\%$. Furthermore, just based on the numerical data, we find another ansatz which fits the results even better
\begin{equation}\label{CLnew}
\frac{1}{2}\sum_{s=\pm1}\frac{g^2\mu^2(1+sh)^2T^2}{\pi^2}\left(c_1^L-c_2^L\ln\frac{g^2}{1+sh}\right),
\end{equation}
where the mixture effect is shown implicitly in the logarithm, since the denominator becomes dimensionless by canceling with the average chemical potential $\mu=\sum_{s=\pm1}\mu(1+sh)/2$ in the numerator.

The specific heat (per volume) at fixed volume and particle number is \cite{Landau:1980-5}
\[C_V=T\left(\frac{\partial S}{\partial T}\right)_V=T\left(\frac{\partial S}{\partial T}\right)_{\mu_f}-\sum_f\frac{[(\partial n_f/\partial T)_{\mu_f}]^2}{(\partial n_f/\partial\mu_f)_T},\]
where the entropy density $S=(\partial\Omega/\partial T)_V$ and particle number density $n_f=-(\partial\Omega/\partial\mu_f)_V$. In the low-temperature limit the second term can be neglected, therefore the specific heat can be obtained, using Eqs.~(\ref{CT}) and (\ref{CL}), as
\begin{widetext}
\begin{equation}\label{heatcapacity}
C_V-C_V^0=-T\left(\frac{\partial^2\Delta\Omega}{\partial T^2}\right)_{\mu_f}=-\frac{1}{2}\sum_{s=\pm1}\frac{2g^2\mu^2(1+sh)^2T}{\pi^2}\left[\frac{1}{6}\ln\frac{4\pi^2(1+sh)^2}{g^2(1+h^2)}-\frac{1}{6}-2c_1^T+3c_2^T+2c_2^T\ln\frac{g^2T}{\mu(1+sh)}\right],
\end{equation}
\end{widetext}
where the specific heat $C_V^0=-T(\partial^2\Delta\Omega_0/\partial T^2)_{\mu_f}$ of an ideal gas has been subtracted.

\begin{figure}
\resizebox{1\linewidth}{!}{\includegraphics{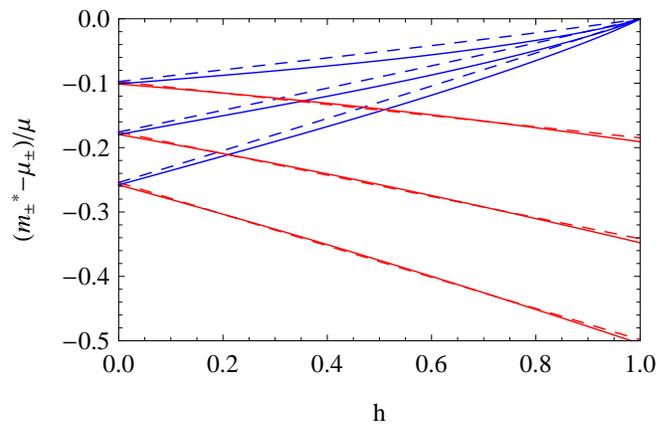}}
\caption{(color online). The RPA correction to the effective mass as function of $h$, with the red curves for the majority flavor ($+$) and the blue curves for the minority flavor ($-$). The solid curves correspond to the expression in Eq.~(\ref{mass}) while the dashed curves to Eq.~(\ref{massnew}), and their difference is not too much. $g=1/2$ and for each group of curves the temperature is $T/\mu=10^{-6}$, $10^{-4}$ and $10^{-2}$ from bottom to top, respectively.\label{fig:mass}}
\end{figure}

The effective mass and the first Landau parameter can be determined in the high-density limit by comparing Eq.~(\ref{CV}) and Eq.~(\ref{heatcapacity}),
\begin{align}\label{mass}
m_\pm^*&=\mu_\pm+\frac{g^2\mu(1\pm h)}{\pi^2}\left[\ln\frac{4\pi^2(1\pm h)^2}{g^2(1+h^2)}-1\right.\nonumber\\
&\left.\qquad\qquad-12c_1^T+18c_2^T+12c_2^T\ln\frac{g^2T}{\mu(1\pm h)}\right],
\end{align}
where we used $k_\pm=\mu_\pm=\mu(1\pm h)$ in the massless limit at high density. The first term in the effective mass is due to the ideal-gas specific heat. However, as pointed out before, the factor $(1+h^2)$ shows the mixing of the two flavors, unlike the weak-coupling results obtained in Sec.~\ref{sec-weakcoupling}, e.g., Eq.~(\ref{imbalancemassexpand}). This is because, with the RPA correction, we incorporate the sum of an infinite chain of gluon self-energies, which incorporates interactions between different quark flavors. Since the longitudinal gluon is screened in the static long-wavelength limit, the mixing is even present in the low-$T$ and small-$g$ limit. The numerical fit in Eq~(\ref{CLnew}) also provides another expression for the effective mass,
\begin{align}\label{massnew}
m_\pm^*&=\mu_\pm+\frac{g^2\mu(1\pm h)}{\pi^2}\left[6c_1^L-6c_2^L\ln\frac{g^2}{1\pm h}\right.\nonumber\\
&\left.\qquad\qquad-12c_1^T+18c_2^T+12c_2^T\ln\frac{g^2T}{\mu(1\pm h)}\right].
\end{align}
As expected, all the above results return to the balanced case as $h\rightarrow0$. As mentioned previously, the contribution from the non-$N_B$ term is not included, which acts as a constant shift on the $c_1^L$ and $c_1^T$ factors. In Fig.~\ref{fig:mass} the change of effective mass due to the interaction is given as function of imbalance at various temperatures. Finally, we emphasize again that $h$ should not be too close to $1$ even though the divergence from $\ln(1-h)$ is suppressed by the prefactor $(1-h)$, because in such an extremely imbalanced case the condition $T\ll\mu_-$ is not satisfied for the minority flavor.

\section{Summary and Discussion}

We have calculated the thermodynamic potential perturbatively in the large-$N_F$ limit and the effective mass of the quarks is determined by using Fermi-liquid theory for an imbalanced cold dense quark system. The temperature dependence is obtained by using the gluon self-energy, from which the contributions from transverse and longitudinal gluons are explicitly shown. For the two-flavor imbalanced quark system, the effective mass is obtained both analytically within the weak-coupling limit, and numerically within the RPA approximation. We find that, in contrast to the weak-coupling result, where the effective mass of each flavor is independent of each other due to the lack of an inter-flavor interaction, the effective mass obtained from the RPA calculation depends on the chemical potentials of both flavors, which is a consequence of the static screening of the electric gluons at long wavelength.

From the RPA results, the logarithmic dependence on temperature of the specific heat and effective mass signals a breakdown of Fermi-liquid theory at zero-temperature. Non-Fermi-liquid behavior, due to unscreened long-range magnetic interactions, was already discussed several decades ago for the case of the electron gas \cite{Holstein:1973,Chakravarty:1995}. In the large-$N_F$ limit the QCD thermodynamic potential is essentially the same as that of QED, apart from group theory factors. The non-abelian effects of QCD only show up if gluon self-interaction corrections are included. The logarithmic behavior in the balanced case has previously been seen in analytic and numerical calculations of the QCD thermodynamic potential and specific heat using the large-$N_F$ limit, dimensional reduction, and hard-dense loop QCD perturbation theory \cite{Ipp:2006,Gerhold:2004,Ipp:2004a,Ipp:2004b,Ipp:2003,Boyanovsky:2001}. We expect that such a logarithmic dependence shall have important effects on the imbalanced QCD phase diagram at low temperatures and directly influence the properties of quark matter in the core of neutron stars. To better understand these effects, it is necessary to go further, such as including the gluon self-interaction and extending our discussion to the three-flavor imbalanced case. We hope progress along these directions will be achieved in the near future.

\section*{Acknowledgments}

This work is supported by the Stichting voor Fundamenteel Onderzoek der Materie (FOM) and the Nederlandse Organisatie voor Wetenschaplijk Onderzoek (NWO).

\appendix

\section{Conventions}

\subsection{Euclidean conventions}\label{app-Euclidean}

The four-momentum vectors in Euclidean space-time will be written
with capital letters $Q=(i(i\omega_n),\mathbf{q})$, while for the
Wick-rotated case ($i\omega_n\rightarrow\omega+i0$) the roman capital
letters $\mathrm{Q}_+=(i(\omega+i0),\mathbf{q})$ are used. Three
momentum vectors are written in bold face $\mathbf{q}$ and its length
as $q$. And $\omega\pm i0$ are sometimes written as $\omega_\pm$ for short.

The Euclidean gamma matrices in the standard representation are
\[
\gamma_0=\left(\begin{array}{cc}
1 & 0\\
0 & -1
\end{array}\right),\quad\gamma_i=-i\left(\begin{array}{cc}
0 & \sigma_{i}\\
-\sigma_i & 0
\end{array}\right),
\]
\[
\gamma_{5}=\gamma_0\gamma_1\gamma_2\gamma_3=\left(\begin{array}{cc}
0 & 1\\
1 & 0
\end{array}\right).
\]
All the above entries are $2\times2$ matrices and $\sigma_i$ are
the Pauli spin matrices
\[
\sigma_1=\left(\begin{array}{cc}
0 & 1\\
1 & 0
\end{array}\right),\quad\sigma_2=\left(\begin{array}{cc}
0 & -i\\
i & 0
\end{array}\right),\quad\sigma_3=\left(\begin{array}{cc}
1 & 0\\
0 & -1
\end{array}\right).
\]
The gamma matrices obey the following relations
\[
\{\gamma_\mu,\gamma_\nu\}=2\delta_{\mu\nu},\quad\{\gamma_5,\gamma_\mu\}=0,\quad\gamma_5^2=1.
\]
The charge conjugation matrix is
\[
C=\gamma_0\gamma_2=-i\left(\begin{array}{cc}
0 & \sigma_2\\
\sigma_2 & 0
\end{array}\right),
\]
which satisfies
\[C\gamma_\mu^TC=\gamma_\mu,\quad C^T=C^{-1}=-C.\]

The eight generators of the fundamental representation of $\mathrm{SU}(3)$
are taken to be
\[
t^1=\frac{1}{2}\left(\begin{array}{ccc}
0 & 1 & 0\\
1 & 0 & 0\\
0 & 0 & 0
\end{array}\right),\quad t^2=\frac{1}{2}\left(\begin{array}{ccc}
0 & -i & 0\\
i & 0 & 0\\
0 & 0 & 0
\end{array}\right),
\]
\[
t^3=\frac{1}{2}\left(\begin{array}{ccc}
1 & 0 & 0\\
0 & -1 & 0\\
0 & 0 & 0
\end{array}\right),\quad t^4=\frac{1}{2}\left(\begin{array}{ccc}
0 & 0 & 1\\
0 & 0 & 0\\
1 & 0 & 0
\end{array}\right),
\]
\[
t^5=\frac{1}{2}\left(\begin{array}{ccc}
0 & 0 & -i\\
0 & 0 & 0\\
i & 0 & 0
\end{array}\right),\quad t^6=\frac{1}{2}\left(\begin{array}{ccc}
0 & 0 & 0\\
0 & 0 & 1\\
0 & 1 & 0
\end{array}\right),
\]
\[
t^7=\frac{1}{2}\left(\begin{array}{ccc}
0 & 0 & 0\\
0 & 0 & -i\\
0 & i & 0
\end{array}\right),\quad t^8=\frac{1}{2\sqrt{3}}\left(\begin{array}{ccc}
1 & 0 & 0\\
0 & 1 & 0\\
0 & 0 & -2
\end{array}\right),
\]
and have been normalized according to $\mathrm{Tr}[t^at^b]=\frac{1}{2}\delta^{ab}$.
In general, the product of the generators is
\[t_{ij}^at_{kl}^a=\frac{N_C-1}{4N_C}(\delta_{ij}\delta_{kl}+\delta_{il}\delta_{kj})-\frac{N_C+1}{4N_C}(\delta_{ij}\delta_{kl}-\delta_{il}\delta_{kj}).\]
Furthermore, one will frequently encounter the following group-theory
factors
\[f^{abc}f^{abd}=N_C\delta^{cd},\quad\delta^{aa}=N_C^2-1\equiv N_G,\]
\[t_{il}^at_{lj}^a=\frac{N_C^2-1}{2N_C}\delta_{ij}=\frac{N_G}{2N_C}\delta_{ij},\]
in the above expressions $N_C$ is the number of colors and $N_G$ the number of gluons.

Fourier transforms are normalized as
\begin{align*}
\psi(\tau,\mathbf{x})&=\frac{1}{\sqrt{\beta}}\sum_n\int\frac{\mathrm{d}^3p}{(2\pi)^{3/2}}\psi(i\omega_n,\mathbf{p})e^{i\mathbf{p}\cdot\mathbf{x}-i\omega_n\tau}\\
&=\int\frac{\mathrm{d}^4p}{\sqrt{\mathcal{V}}}\psi(p)e^{ip_\mu x_\mu},
\end{align*}
where $\omega_n=\pi(2n+1)/\beta$ are the fermionic Matsubara frequencies,
$\mathcal{V}=\beta(2\pi)^3$ is the imaginary time phase space volume
and $p_\mu=(i(i\omega_n),\mathbf{p})$ and $x_\mu=(\tau,\mathbf{x})$.

The Dirac equation in Euclidean space is
\[(\slashed{\partial}+m)\psi(x)=0,\]
where $\slashed{\partial}=\gamma_\mu\partial_\mu$. In momentum
space the Dirac equation reads
\[(i\slashed{p}+m)\psi(p)=0.\]
The eigenvalues of the matrix $i\slashed{p}$ are $\pm m$, since
$(i\slashed{p})^2=-p^2=m^2$. The eigenspinors corresponding to these
eigenvalues are
\begin{align*}
i\slashed{p}u_s(\mathbf{p})&=-mu_s(\mathbf{p}),\\
i\slashed{p}v_s(\mathbf{p})&=mv_s(\mathbf{p}).
\end{align*}
Note that $i\slashed{p}$ is not a hermitian matrix and that $v_s(-\mathbf{p})$
satisfies the same equation as $u_s(\mathbf{p})$. However,
they can also be viewed as the the eigenspinors of a hermitian matrix
\begin{align*}
\gamma_0(i\mathbf{p}\cdot\vec{\gamma}+m)u_s(\mathbf{p})&=\epsilon_pu_s(\mathbf{p}),\\
\gamma_0(i\mathbf{p}\cdot\vec{\gamma}+m)v_s(-\mathbf{p})&=-\epsilon_pv_s(-\mathbf{p}).
\end{align*}
The positive and negative energy eigenspinors have the following form
\begin{align}
u_s(\mathbf{p})&=\sqrt{\frac{\epsilon_p+m}{2\epsilon_p}}\left(\begin{array}{c}
\xi_s\\
\frac{\mathbf{p}\cdot\vec{\sigma}}{\epsilon_p+m}\xi_s
\end{array}\right),\label{eq:AP_Eigenspinors}\\
v_s(-\mathbf{p})&=\sqrt{\frac{\epsilon_p+m}{2\epsilon_p}}\left(\begin{array}{c}
-\frac{\mathbf{p}\cdot\vec{\sigma}}{\epsilon_p+m}\xi_s\\
\xi_s
\end{array}\right),
\end{align}
where $\xi_{\uparrow}=(1,0)^T$ and $\xi_{\downarrow}=(0,1)^T$. They
are orthonormal in the sense that
\begin{align*}
u_s^{\dagger}(\mathbf{p})u_s'(\mathbf{p})&=v_s^{\dagger}(-\mathbf{p})v_s'(-\mathbf{p})=\delta_{ss'},\\
v_s^{\dagger}(-\mathbf{p})u_s'(\mathbf{p})&=u_s^{\dagger}(-\mathbf{p})v_s'(\mathbf{p})=0,
\end{align*}
and satisfy the completeness relation
\[\sum_s[u_s(\mathbf{p})u_s^\dagger(\mathbf{p})+v_s(-\mathbf{p})v_s^\dagger(-\mathbf{p})]=1.\]

Using the above eigenspinors the positive and negative energy projectors
can be defined as
\begin{align*}
P_E^+(\mathbf{p})&\equiv\sum_su_s(\mathbf{p})u_s^\dagger(\mathbf{p})=\frac{\epsilon_p\gamma_0-i\vec{\gamma}\cdot\mathbf{p}+m}{2\epsilon_p}\gamma_0,\\
P_E^-(\mathbf{p})&\equiv\sum_sv_s(-\mathbf{p})v_s^\dagger(-\mathbf{p})=\frac{\epsilon_p\gamma_0+i\vec{\gamma}\cdot\mathbf{p}-m}{2\epsilon_p}\gamma_0,
\end{align*}
which in the massless case reduce to
\begin{align*}
P_{E}^{s}(\mathbf{p},m=0) & =\frac{1+si\gamma_{0}\vec{\gamma}\cdot\hat{\mathbf{p}}}{2}.
\end{align*}
The helicity projection operators project the spin along the momentum
of the particle and read
\[\mathcal{P}^s(\mathbf{p})=\frac{1+s\mathbf{\Sigma}\cdot\hat{\mathbf{p}}}{2}=\frac{1+si\gamma_5\gamma_0\vec{\gamma}\cdot\hat{\mathbf{p}}}{2},\]
where $\mathbf{\Sigma}=i\gamma_5\gamma_0\vec{\gamma}=\left(\begin{array}{cc}
\vec{\sigma} & 0\\
0 & \vec{\sigma}
\end{array}\right)$ is the spin operator. Note that $[P_E^s(\mathbf{p})\gamma_0,\mathcal{P}^{s'}(\mathbf{p})]=0$.

\subsection{The Green's functions}\label{app-Green}

The Green's functions of quarks and gluons in Euclidean space are defined by
\begin{align*}
G(\tau,\mathbf{x};\tau',\mathbf{x}')&=-\langle\psi(\tau,\mathbf{x})\bar{\psi}(\tau',\mathbf{x}')\rangle,\\
D_{\mu\nu}(\tau,\mathbf{x};\tau',\mathbf{x}')&=\langle A_\mu(\tau,\mathbf{x})A_\nu(\tau',\mathbf{x}')\rangle.
\end{align*}
In momentum space, the free quark propagator is
\[G_0(p)=\frac{(i\omega_n+\mu)\gamma_0-i\vec{\gamma}\cdot\mathbf{p}+m}{(i\omega_n+\mu)^2-p^2-m^2}=\frac{i\slashed{P}-m}{P^2+m^2}.\]
In the Lorentz gauge $(\partial_\mu A_\mu=0)$, the momentum-space free gluon propagator is
\[D_{0,\mu\nu}(q)=\frac{1}{Q^2}\left[\delta_{\mu\nu}-(1-\xi)\frac{Q_\mu Q_\nu}{Q^2}\right]=\frac{P_{\mu\nu}}{Q^2}+\frac{\xi}{Q^2}\frac{Q_\mu Q_\nu}{Q^2},\]
while in the Coulomb gauge $\left(\partial_iA_i=0\right)$ it
is \cite{Le Bellac:1996}
\[D_{0,\mu\nu}=\frac{P_{\mu\nu}^T}{Q^2}+\frac{Q^2}{q^2}\frac{\delta_{\mu0}\delta_{\nu0}}{Q^2}+\xi\frac{Q^2}{q^4}\frac{Q_\mu Q_\nu}{Q^2},\]
where $\xi$ is a gauge-fixing parameter and the projectors can be
written as
\begin{align*}
P_{\mu\nu}&=\delta_{\mu\nu}-\frac{Q_{\mu}Q_{\nu}}{Q^2},\\
P_{ij}^T&=\delta_{ij}-\frac{\mathbf{q}_i\mathbf{q}_j}{q^2},\quad P_{\mu0}^T=P_{0\nu}^T=0,\\
P_{\mu\nu}^L&=P_{\mu\nu}-P_{\mu\nu}^T.
\end{align*}
Note that the free gluon propagator in the Lorentz gauge satisfies
$Q_\mu D_{0,\mu\nu}=\xi Q_\nu/Q^2$, such that in the Landau
gauge ($\xi=0$) it is purely four-momentum transverse, i.e., $Q_\mu D_{0,\mu\nu}=0$.
The free gluon propagator in the Coulomb gauge satisfies $\mathbf{q}_iD_{0,i\nu}=\xi q^2Q_\nu/Q^4$,
such that in the Landau gauge it is three-momentum transverse. These
propagators are diagonal in color space.

\section{Lorentz transformation properties}\label{app-Lorentz}

In this section the Lorentz transformation properties of some quantities
are summarized, such as the thermodynamic potential, the distribution
function and the effective interaction. The Lorentz transformation
to a frame moving with velocity $\mathbf{v}$ is
\[
\Lambda^{\mu\nu}(\mathbf{v})=\left(\begin{array}{cccc}
-\gamma & \cdots & -\gamma\mathbf{v}^{T} & \cdots\\
\vdots & \ddots\\
\gamma\mathbf{v} &  & \delta_{ij}+\frac{\mathbf{v}_{i}\mathbf{v}_{j}}{v^2}(\gamma-1)\\
\vdots &  &  & \ddots
\end{array}\right),
\]
such that the four-momentum $P^{\mu}=(\epsilon(\mathbf{p}),\mathbf{p})$
of a particle transforms as
\[
\left(\begin{array}{c}
\epsilon(\mathbf{p})\\
\mathbf{p}
\end{array}\right)\rightarrow\left(\begin{array}{c}
\gamma(\epsilon(\mathbf{p})-\mathbf{v}\cdot\mathbf{p})\\
\mathbf{p}+\hat{\mathbf{v}}(\mathbf{p}\cdot\hat{\mathbf{v}})(\gamma-1)-\gamma\epsilon(\mathbf{p})\mathbf{v}
\end{array}\right).
\]
Not to be confused with the Dirac matrices $\gamma^\mu$, the $\gamma$ used in this section is the Lorentz factor $\gamma=1/\sqrt{1-v^2}$.

\subsection{Lorentz invariance of the thermodynamic potential density}\label{app-invariance}

The invariance of the thermodynamic potential density under Lorentz transformations
can be shown using the stress-energy tensor. Consider the change in
the thermodynamic potential density under a Lorentz transformation
from the rest frame to a frame moving with velocity $\mathbf{v}$.
Since an interacting gas of quarks in the rest frame is specified
only by an energy density $\rho$ and a pressure $p$, the stress-energy
tensor is diagonal and of the form
\[T_{\mu\nu}=\mathrm{diag}(\rho,p,p,p)_{\mu\nu},\]
such that under a Lorentz transformation the thermodynamic potential density
$\Omega=\rho-\mu n=-p$ changes as
\begin{align*}
\delta\Omega&=\delta T_{00}-\delta(\mu n)=\gamma^2v^2(\rho+p)-\delta(\mu n)\\
&=\mu n\gamma^2v^2-\delta(\mu n).
\end{align*}
The transformation of the chemical potential ($\mu=\partial E/\partial N$)
follows from the Lorentz boosted total energy of the system $E\rightarrow\gamma(E-\mathbf{v}\cdot\mathbf{P})=\gamma E$,
where $\mathbf{P}=\mathbf{0}$ is the total momentum of the system
in the rest frame, which results in $\mu\rightarrow\gamma\mu$. The
transformation of the density is due to a Lorentz contraction in the
volume $n\rightarrow\gamma n$. Thus the total change in $\mu n$
is $\delta(\mu n)=(\gamma^2-1)\mu n=\gamma^2v^2\mu n$,
such that under a Lorentz transformation $\delta\Omega=0$. For an
alternative derivation which uses that the pressure transforms the
same way as a force per area, see Ref. \cite{Tolman:1987}.

\subsection{Transformation properties of $f(\mathbf{p},\mathbf{p}')$}\label{app-transformation}

Consider the lowest-order correction to the free thermodynamic potential density
\[\Omega_{\mathrm{int}}=\frac{1}{2}\sum_{\sigma,\sigma'}\int\frac{\mathrm{d}^3p\mathrm{d}^3p'}{(2\pi)^{3}}f_{\sigma\sigma'}(\mathbf{p},\mathbf{p}')N_\sigma(\mathbf{p})N_{\sigma'}(\mathbf{p}').\]
Note that the distribution function $N(\mathbf{p})$ is a Lorentz
invariant, which can be easily derived from the fact that the number
of particles in a volume $\mathrm{d}^3x\mathrm{d}^3p$
of phase space is invariant under Lorentz transformations \cite{Landau:1980-2},
i.e., $\tilde{N}(\tilde{\mathbf{p}})=N(\mathbf{p})$ where the tilde
signifies the Lorentz transformed quantity. Subsequently, it is possible
to derive a transformation law for $f(\mathbf{p},\mathbf{p}')$ by
using that the distribution, the thermodynamic potential density and $\mathrm{d}\mathbf{p}/\epsilon_0(\mathbf{p})$ are Lorentz invariant. It follows that $\epsilon_{0\sigma}(\mathbf{p})\epsilon_{0\sigma'}(\mathbf{p}')f_{\sigma\sigma'}(\mathbf{p},\mathbf{p}')$
should be Lorentz invariant, giving
\[\epsilon_{0\sigma}(\mathbf{p})\epsilon_{0\sigma'}(\mathbf{p}')f_{\sigma\sigma'}(\mathbf{p},\mathbf{p}')=\tilde{\epsilon}_{0\sigma}(\tilde{\mathbf{p}})\tilde{\epsilon}_{0\sigma'}(\tilde{\mathbf{p}}')\tilde{f}_{\sigma\sigma'}(\tilde{\mathbf{p}},\tilde{\mathbf{p}}'),\]
where $\epsilon_{0\sigma}(\mathbf{p})=\sqrt{p^2+m_\sigma^2}$. Expand $\tilde{f}$ to the lowest order in $\mathbf{v}$, using $\tilde{\epsilon}_0(\tilde{\mathbf{p}})=\epsilon_0(\mathbf{p})-\mathbf{v}\cdot\mathbf{p}+\mathcal{O}(v^2)$, $\tilde{\mathbf{p}}=\mathbf{p}-\epsilon_0(\mathbf{p})\mathbf{v}+\mathcal{O}(v^2)$, and $\mathbf{p}/\epsilon_0(\mathbf{p})=\partial\epsilon_0(\mathbf{p})/\partial\mathbf{p}$,
\begin{widetext}
\begin{align*}
&f_{\sigma\sigma'}(\mathbf{p},\mathbf{p}')=\tilde{f}_{\sigma\sigma'}(\tilde{\mathbf{p}},\tilde{\mathbf{p}}')\left[1-\mathbf{v}\cdot\frac{\mathbf{p}}{\epsilon_{0\sigma}(\mathbf{p})}\right]\left[1-\mathbf{v}\cdot\frac{\mathbf{p}'}{\epsilon_{0\sigma'}(\mathbf{p}')}\right]+\mathcal{O}(v^2)\\
=&\tilde{f}_{\sigma\sigma'}(\mathbf{p},\mathbf{p}')-\epsilon_{0\sigma}(\mathbf{p})\mathbf{v}\cdot\frac{\partial f_{\sigma\sigma'}(\mathbf{p},\mathbf{p}')}{\partial\mathbf{p}}-\epsilon_{0\sigma'}(\mathbf{p}')\mathbf{v}\cdot\frac{\partial f_{\sigma\sigma'}(\mathbf{p},\mathbf{p}')}{\partial\mathbf{p}'}-f_{\sigma\sigma'}(\mathbf{p},\mathbf{p}')\left[\mathbf{v}\cdot\frac{\partial\epsilon_{0\sigma}(\mathbf{p})}{\partial\mathbf{p}}-\mathbf{v}\cdot\frac{\partial\epsilon_{0\sigma'}(\mathbf{p}')}{\partial\mathbf{p}'}\right]+\mathcal{O}(v^2)\\
=&\tilde{f}_{\sigma\sigma'}(\mathbf{p},\mathbf{p}')-\mathbf{v}\cdot\left[\frac{\partial\epsilon_{0\sigma}(\mathbf{p})f_{\sigma\sigma'}(\mathbf{p},\mathbf{p}')}{\partial\mathbf{p}}+\frac{\partial\epsilon_{0\sigma'}(\mathbf{p}')f_{\sigma\sigma'}(\mathbf{p},\mathbf{p}')}{\partial\mathbf{p}'}\right]+\mathcal{O}(v^2).
\end{align*}
\end{widetext}
If it is assumed that to the lowest order the interaction does not depend
on any distribution functions, i.e., $\tilde{f}_{\sigma\sigma'}(\mathbf{p},\mathbf{p}')=f_{\sigma\sigma'}(\mathbf{p},\mathbf{p}')$,
the above implies
\begin{equation}
\frac{\partial\epsilon_{0\sigma}(\mathbf{p})f_{\sigma\sigma'}(\mathbf{p},\mathbf{p}')}{\partial\mathbf{p}}=-\frac{\partial\epsilon_{0\sigma'}(\mathbf{p}')f_{\sigma\sigma'}(\mathbf{p},\mathbf{p}')}{\partial\mathbf{p}'}.\label{eq:AP_eff_int_transf}
\end{equation}
Note that the above derivation is similar to that given in Ref.~\cite{Baym:1976}.

\section{From Minkowski to Euclidian space}\label{app-MinToEuc}

In this section it is summarized how to turn the Minkowski quantum
field theory (QFT) of quantum chromodynamics into a Euclidean statistical
field theory (SFT) suitable for studying the dynamical properties
of a many-particle system. The starting point is the gauge-fixed path
integral for QCD
\begin{equation}
\int\mathscr{D}A_{\mu}\mathscr{D}\bar{\psi}\mathscr{D}\psi\mathscr{D}\bar{\eta}\mathscr{D}\eta\:\exp\left\{ i\int\mathscr{L}_{\textrm{QCD}}\textrm{d}^{4}x\right\} ,\label{eq:AP_Minkowski_path_int}
\end{equation}
where $A_{\mu}$ are the gluon fields, $\psi$ and $\bar{\psi}=i\psi^{\dagger}\gamma^0$
the quark fields, $\eta$ and $\bar{\eta}=i\eta^{\dagger}\gamma^0$
the ghost fields and the QCD Lagrangian in Minkowski space fixed in
a linear gauge ($f^\mu A_\mu^a=0$) is given by
\begin{align}\label{eq:AP_Minkowski_QCD_Lagrangian}
\mathscr{L}_{QCD}=&-\sum_f[\bar{\psi}_f(\slashed{\partial}+m_f)\psi_f+ig\bar{\psi}_f\gamma^\mu t^a\psi_fA_\mu^a]\nonumber\\
&-\frac{1}{2}A_a^\mu(\partial_{\mu}\partial_\nu-\partial^2\eta_{\mu\nu})A_a^\nu\nonumber\\
&-\bar{\eta}^a\partial_\mu\partial^\mu\eta^a-gf^{abc}(\bar{\eta}^a\partial^\mu\eta^b)A_\mu^c\nonumber\\
&-\frac{1}{2}g(\partial^\mu A_a^\nu-\partial^\nu A_a^\mu)f^{abc}\eta_{\mu\sigma}\eta_{\nu\rho}A_b^\sigma A_c^\rho\nonumber\\
&-\frac{1}{4}g^2f^{abc}f^{ade}\eta_{\mu\sigma}\eta_{\nu\rho}A_a^\mu A_c^\nu A_d^\sigma A_e^\rho+\frac{1}{2\xi}(f^\mu A_\mu^a)^2,
\end{align}
where $\psi_f$ is the quark field with flavor $f$, and the summation over color and spin indices are shown implicitly. The metric was chosen to be $\eta_{\mu\nu}=\mathrm{diag}\left(-1,1,1,1\right)$
and the Minkowski gamma matrices in the standard representation are
\begin{align}
\gamma^0&=-i\left(\begin{array}{cc}
1 & 0\\
0 & -1
\end{array}\right),\quad\gamma^i=-i\left(\begin{array}{cc}
0 & \sigma_i\nonumber\\
-\sigma_i & 0
\end{array}\right),\\
\gamma^5&=i\gamma^0\gamma^1\gamma^2\gamma^3=\left(\begin{array}{cc}
0 & 1\\
1 & 0
\end{array}\right).\label{eq:AP_Minkowski_gamma}
\end{align}
The gamma matrices satisfy
\[\{\gamma^\mu,\gamma^\nu\}=2\eta^{\mu\nu},\quad\{\gamma^5,\gamma^\mu\}=0,\quad(\gamma^5)^2=1.\]

By performing a Wick rotation the above quantum field theory can be
transformed into a statistical field theory. A Wick rotation amounts
to taking an analytic continuation from real time to imaginary time
($t=-i\tau$), which turns the Minkowski metric $\mathrm{ds}^2=-\mathrm{d}t^2+\mathrm{d}x^2$
into the Euclidean metric $\mathrm{ds}^2=\mathrm{d}\tau^2+\mathrm{d}x^2$.
To do this consistently the zeroth component of all four-vectors need
to change accordingly. The procedure is most easily understood by
considering the length of the position four-vector
\begin{align*}
x^\mu x_\mu=x^\mu\eta_{\mu\nu}x^\nu& =(t,\mathbf{x})\left(\begin{array}{cc}
-1 & 0\\
0 & \mathbf{1}
\end{array}\right)\left(\begin{array}{c}
t\\
\mathbf{x}
\end{array}\right)\\
&=(it,\mathbf{x})\ \mathbb{I}\left(\begin{array}{c}
it\\
\mathbf{x}
\end{array}\right)\equiv x_\mu^E\delta_{\mu\nu}x_\nu^E,\\
x_\mu\eta^{\mu\nu}x_\nu&=(-t,\mathbf{x})\left(\begin{array}{cc}
-1 & 0\\
0 & \mathbf{1}
\end{array}\right)\left(\begin{array}{c}
-t\\
\mathbf{x}
\end{array}\right)\\
&=(it,\mathbf{x})\ \mathbb{I}\left(\begin{array}{c}
it\\
\mathbf{x}
\end{array}\right)\equiv x_\mu^E\delta_{\mu\nu}x_\nu^E,
\end{align*}
where it is seen that the minus sign of the Minkowski metric is absorbed
in the definition of the Euclidean four-vectors $x_\mu^E=(it,\mathbf{x})=(\tau,\mathbf{x})$.
In Euclidean space no distinction is made between upper and lower
indices. A simple way to obtain the Euclidean form of a vector is
to multiply the contravariant vector by the matrix $\mathrm{diag}\left(i,1,1,1\right)$
or the covariant vector by $\mathrm{diag}\left(-i,1,1,1\right)$ and
set $t=-i\tau$. Note that the spatial components do not change. For
example, the Euclidean position four-vector, the four-divergence and
the zeroth gamma matrix are in terms of their Minkowski definitions
Eq.~(\ref{eq:AP_Minkowski_gamma})
\begin{align*}
x_\mu^E&=(ix^0,\mathbf{x})=(-ix_0,\mathbf{x})=(it,\mathbf{x})=(\tau,\mathbf{x}),\\
\partial_\mu^E&=(i\partial^0,\nabla)=(-i\partial_0,\nabla)=\left(-i\frac{\partial}{\partial t},\nabla\right)=(\partial_\tau,\nabla),\\
\gamma_0^E&=i\gamma^0=-i\gamma_0=\left(\begin{array}{cc}
1 & 0\\
0 & -1
\end{array}\right).
\end{align*}

Generalizations to tensors is straightforward and follows for instance from the example $A_{\mu\nu}=a_\mu a_\nu$. Using the above procedure to find the Euclidean versions of all tensors, the partition function is easily found from Eqs.~(\ref{eq:AP_Minkowski_path_int}) and (\ref{eq:AP_Minkowski_QCD_Lagrangian}) by setting the tensors to their Euclidean versions, taking $\eta_{\mu\nu}\rightarrow\delta_{\mu\nu}$ and $t\rightarrow-i\tau$. The partition function is
\[Z=\int\mathscr{D}A_\mu\mathscr{D}\bar{\psi}\mathscr{D}\psi\mathscr{D}\bar{\eta}\mathscr{D}\eta\: e^{-S^{\mathrm{E}}},\]
where the Euclidean action is defined as
\[S^{\mathrm{E}}=\int\mathscr{L}_{\mathrm{QCD}}^{\mathrm{E}}\textrm{d}\tau\textrm{d}\mathbf{x},\]
with the Euclidean Lagrangian
\begin{align*}
\mathscr{L}_{\textrm{QCD}}^E&=\sum_f[\bar{\psi}_f(\slashed{\partial}+m_f-\gamma_0\mu_f)\psi_f-ig\bar{\psi}_f\gamma_\mu t^a\psi_fA_\mu^a]\\
&+\frac{1}{2}A_\mu^a(\partial_\mu\partial_\nu-\partial^2\delta_{\mu\nu})A_\nu^a\\
&+\bar{\eta}^a\partial^2\eta^a+gf^{abc}(\bar{\eta}^a\partial_\mu\eta^b)A_\mu^c\\
&+\frac{1}{2}g(\partial_\mu A_\nu^a-\partial_\nu A_\mu^a)f^{abc}A_\mu^bA_\nu^c\\
&+\frac{1}{4}g^2f^{abc}f^{ade}A_\mu^bA_\nu^cA_\mu^dA_\nu^e-\frac{1}{2\xi}(\partial_\mu A_\mu^a)^2.
\end{align*}
In the above the tensors are all Euclidean but the index $E$ has
been dropped for convenience, the conjugate fields are now defined
as $\bar{\psi}_f=\psi_f^\dagger\gamma_0^E$ and $\bar{\eta}=\eta^\dagger\gamma_0^E$
and the chemical potential has been added as the Lagrange multiplier
of the density $\psi_f^\dagger\psi_f$. Additionally, to complete the
connection between QFT and SFT, the time integration domain is changed
to $\tau\in[0,\beta]$, where $\beta$ is the inverse temperature
$T$ of the system. Due to the definition of the partition function
as a trace over all states, the bosonic (fermionic) fields are required
to obey symmetric (anti-symmetric) boundary conditions, namely $\psi_f(\tau=0,\mathbf{x})=\pm\psi_f(\tau=\beta,\mathbf{x})$.

\section{Nonzero temperature calculations}\label{app-Matsubara}

\begin{figure}
\includegraphics[scale=0.4]{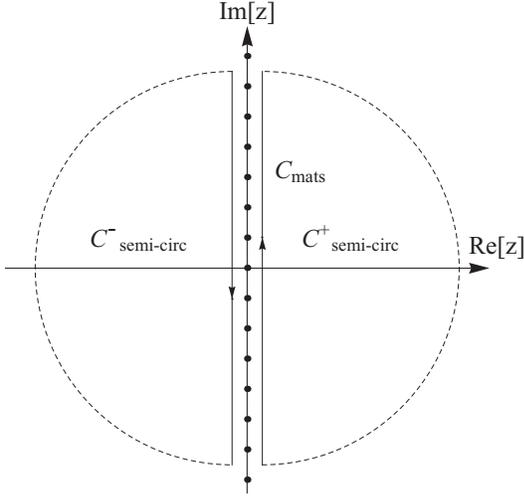}
\caption{The contour for the bosonic Matsubara summation, where $C_\mathrm{mats}$ along the imaginary axis is used to rewrite the frequency summation as a contour integral. The contours $C_\mathrm{semi-circ}^\pm$ are the positive and negative energy semicircular contours obtained by adding arcs at infinity to $C_\mathrm{mats}$.\label{fig:AP_bose_matsubara_contours}}
\end{figure}

At nonzero temperature one needs to calculate Matsubara summations
which usually can be done using contour integration. In the following
an expression is derived for such summations and the interpretation
of the result is discussed. Consider to this end the sum over bosonic
Matsubara frequencies ($\omega_n=2n\pi T$) of the function $f(i\omega_n)$
\[\beta^{-1}\sum_{\omega_n}f(i\omega_n)=\frac{1}{2\pi i}\ointop_{C_\mathrm{mats}}f(z)\frac{1}{e^{\beta z}-1}\mathrm{d}z,\]
where the contour $C_\mathrm{mats}$ is given in Fig.~\ref{fig:AP_bose_matsubara_contours}.
If $f(z)N_B(z)\rightarrow0$ when $|z|\rightarrow\infty$ then it is possible to close the contour by adding the arcs of $C_\mathrm{semi-circ}^\pm$, in which case $C_\mathrm{mats}=C_\mathrm{semi-circ}^++C_\mathrm{semi-circ}^-$.
In general $f(z)$ will only have poles or branch cuts on the real axis, such that the contours can be contracted along the real axis, which gives
\begin{align}
\beta^{-1}\sum_{\omega_n}f(i\omega_n)=&\frac{1}{\pi}\int_0^\infty N_B(\omega)\Im[f(\omega+i0)]\mathrm{d}\omega\nonumber\\
&-\frac{1}{\pi}\int_0^\infty[1+N_B(\omega)]\Im[f(-\omega+i0)]\mathrm{d}\omega,\label{eq:AP_matsubara_bosons}
\end{align}
where $2i\Im[f(\omega+i0)]\equiv f(\omega+i0)-f(\omega-i0)$ and it
was used that $\omega<0$ for the contour $C_\mathrm{semi-arc}^-$,
such that it is more convenient to take $\omega\rightarrow-\omega$ and
use $N_B(-\omega)=-1-N_B(\omega)$ . The first line can be interpreted
as due to thermal gluons and the second line due to thermal (anti)gluons
and a vacuum contribution. The fact that the gluon is its own antiparticle
will be reflected by $\Im[f(-\omega+i0)]=-\Im[f(\omega+i0)]$.

Let us evaluate the above for the specific case $f(z)=\ln[g(z)]$ which satisfies $\Im[g(-\omega+i0)]=-\Im[g(\omega+i0)]$. Expanding the logarithm in terms of its real and imaginary parts
\begin{align*}
\ln[g(z)]=&\ln|g(z)|+i\arctan\left(\frac{\Im[g(z)]}{\Re[g(z)]}\right)\\
&+i\pi\Theta(-\Re[g(z)])\mathrm{sgn}(\Im[g(z)]),
\end{align*}
then Eq.~(\ref{eq:AP_matsubara_bosons}) can be written as
\begin{align*}
&\beta^{-1}\sum_{\omega_n}\ln[g(i\omega_n)]\\
=&\frac{1}{\pi}\int_0^\infty\mathrm{d}\omega(1+2N_{B}(\omega))\left[\arctan\left(\frac{\Im\left[g(\omega_+)\right]}{\Re\left[g(\omega_+)\right]}\right)\right.\\
&\hphantom{\times}\left.+\pi\Theta(-\Re[g(\omega_+)])\mathrm{sgn}(\Im[g(\omega_+)])\vphantom{\arctan\left(\frac{\Im\left[g(\omega_+)\right]}{\Re\left[g(\omega_+)\right]}\right)}\right],
\end{align*}
which will be used in the calculation of the RPA correction to the thermodynamic potential.

\begin{figure}
\includegraphics[scale=0.41]{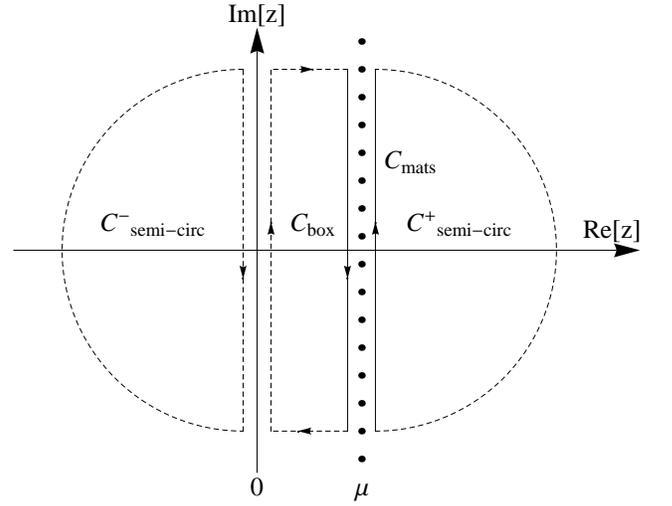}
\caption{The same as in Fig.~\ref{fig:AP_bose_matsubara_contours} but for the fermionic case, where $C_\mathrm{box}$, the contour between $0<\omega<\mu$, is related to processes inside the Fermi sphere.\label{fig:AP_fermi_matsubara_contours}}
\end{figure}

For fermionic Matsubara frequencies a similar derivation can be done,
but now the chemical potential is included by writing $f(i\omega_n+\mu)$
\[\beta^{-1}\sum_{\omega_n}f(i\omega_n+\mu)=-\frac{1}{2\pi i}\ointop_{C_\mathrm{mats}}f(z)\frac{1}{e^{\beta(z-\mu)}+1}\mathrm{d}z,\]
where the contour $C_\mathrm{mats}$ is given in Fig.~\ref{fig:AP_fermi_matsubara_contours}
and $N(z)\equiv(\exp(\beta z)+1)^{-1}$ has poles at the
fermionic Matsubara frequencies with residue $-1$. If $f(z)N(z)\rightarrow0$
when $|z|\rightarrow\infty$, then it is possible to close
the contour as given in Fig.~\ref{fig:AP_fermi_matsubara_contours} by
adding the arcs of $C_\mathrm{semi-circ}^\pm$ and the lower-
and upper-boundaries of $C_\mathrm{box}$. The full Matsubara sum
can thus be written as a contour integral over $C_{\mathrm{mats}}=C_{\mathrm{semi-circ}}^{+}+C_{\mathrm{box}}+C_{\mathrm{semi-circ}}^{-}$.
Again if $f(z)$ only has poles and branch cuts on the real axis,
contracting the contours along the real axis gives
\begin{align}\label{eq:AP_matsubara_fermions}
\beta^{-1}&\sum_{\omega_n}f(i\omega_n+\mu)\nonumber\\
=&-\frac{1}{\pi}\int_\mu^\infty N(\omega-\mu)\Im[f(\omega+i0)]\mathrm{d}\omega\nonumber\\
&-\frac{1}{\pi}\int_0^\mu[1-N(\mu-\omega)]\Im[f(\omega+i0)]\mathrm{d}\omega\nonumber\\
&-\frac{1}{\pi}\int_0^\infty[1-N(\omega+\mu)]\Im[f(-\omega+i0)]\mathrm{d}\omega.
\end{align}
where in the first-to-last line it was used that $0<\omega<\mu$ for the
contour $C_{\mathrm{box}}$, such that it is more convenient to write
$N(\omega-\mu)=1-N(\mu-\omega)$. Similarly for $C_{\mathrm{semi-circ}}^{-}$,
where $\omega<0$, we take $\omega\rightarrow-\omega$ in the integral
and write $N(-\omega-\mu)=1-N(\omega+\mu)$. From the above it is
clear that the first line corresponds to particles above the Fermi
sphere, the second line to those in the Fermi sphere and the last
line is due to anti-particles and contains a vacuum contribution.
Note that the above formula must be used with care, since one has
to check if the above simplifications are valid on a case-to-case
basis.

\section{The gluon self-energy}\label{app-gluon-selfenergy}

The gluon self-energy, or the so-called polarization tensor, is given
to the lowest order by
\begin{align}\label{eq:def_Polarization_tensor}
\Pi_{f,\mu\nu}^{ab}&=-g^2t_{ij}^at_{ji}^b\int\frac{\mathrm{d}^4P}{V}\mathrm{Tr}_s[\gamma_{\mu}G_{0f}(P+Q)\gamma_{\nu}G_{0f}(P)]\nonumber\\
&\equiv\delta^{ab}\Pi_{f,\mu\nu}.
\end{align}
The polarization tensor is purely transverse, i.e., $Q_{\mu}\Pi_{f,\mu\nu}=0$,
which can be easily seen using
\[i\slashed{Q}=G_{0f}^{-1}(P+Q)-G_{0f}^{-1}(P),\]
and the cyclicity of the trace. Expanding the polarization tensor
in terms of the longitudinal and transverse parts relative to the
three-momentum $\mathbf{q}$ gives
\[\Pi_{f,\mu\nu}=F_fP_{\mu\nu}^L+G_fP_{\mu\nu}^T,\]
where $F_f$ and $G_f$ can be obtained by
\begin{equation}
F_f=\frac{Q^2}{q^2}\Pi_{f,00},\quad G_f=\frac{1}{2}\left(\Pi_{f,\mu\mu}-F_f\right).\label{FG}
\end{equation}
Noticing that even in the imbalanced case, the contribution from each flavor simply adds up to the total polarization tensor, therefore, throughout this section, we will not explicitly write out the subscript $f$, and all the formulae presented here are applicable to any flavor. However, in the main text, we use $\Pi_{\mu\nu}$, $F$ and $G$ as the summation over all flavors if the subscript $f$ is not explicitly shown.

After performing the Matsubara sum, the individual processes involved
can be identified
\begin{widetext}
\begin{align}\label{PolarizationTensor}
\Pi_{\mu\nu}=&\frac{g^2}{2}\int\frac{\mathrm{d}^3p}{(2\pi)^3}\left\{\frac{\sum_{ss'}[\bar{u}_s(\mathbf{p})\gamma_{\mu}u_{s'}(\mathbf{p}+\mathbf{q})\bar{u}_{s'}(\mathbf{p}+\mathbf{q})\gamma_{\nu}u_s(\mathbf{p})]}{i\omega_q-\epsilon_0(\mathbf{p}+\mathbf{q})+\epsilon_0(\mathbf{p})}[N(\mathbf{p}+\mathbf{q})(1-N(\mathbf{p}))-(1-N(\mathbf{p}+\mathbf{q}))N(\mathbf{p})]\right.\nonumber\\
&\qquad\qquad\quad\left.+\frac{\sum_{ss'}[\bar{v}_s(\mathbf{p})\gamma_{\mu}u_{s'}(\mathbf{p}+\mathbf{q})\bar{u}_{s'}(\mathbf{p}+\mathbf{q})\gamma_{\nu}v_s(\mathbf{p})]}{i\omega_q-\epsilon_0(\mathbf{p}+\mathbf{q})-\epsilon_0(\mathbf{p})}[N(\mathbf{p}+\mathbf{q})\bar{N}(\mathbf{p})-(1-N(\mathbf{p}+\mathbf{q}))(1-\bar{N}(\mathbf{p}))]\right.\nonumber\\
&\qquad\qquad\quad\left.+\frac{\sum_{ss'}[\bar{u}_s(\mathbf{p})\gamma_{\mu}v_{s'}(\mathbf{p}+\mathbf{q})\bar{v}_{s'}(\mathbf{p}+\mathbf{q})\gamma_{\nu}u_s(\mathbf{p})]}{i\omega_q+\epsilon_0(\mathbf{p}+\mathbf{q})+\epsilon_0(\mathbf{p})}[(1-\bar{N}(\mathbf{p}+\mathbf{q}))(1-N(\mathbf{p}))-\bar{N}(\mathbf{p}+\mathbf{q})N(\mathbf{p})]\right.\nonumber\\
&\qquad\qquad\quad\left.+\frac{\sum_{ss'}[\bar{v}_s(\mathbf{p})\gamma_{\mu}v_{s'}(\mathbf{p}+\mathbf{q})\bar{v}_{s'}(\mathbf{p}+\mathbf{q})\gamma_{\nu}v_s(\mathbf{p})]}{i\omega_q+\epsilon_0(\mathbf{p}+\mathbf{q})-\epsilon_0(\mathbf{p})}[(1-\bar{N}(\mathbf{p}+\mathbf{q}))\bar{N}(\mathbf{p})-\bar{N}(\mathbf{p}+\mathbf{q})(1-\bar{N}(\mathbf{p}))]\right\}\nonumber\\
=&\frac{g^2}{2}\sum_{s_1,s_2=\pm1}\int\frac{\mathrm{d}^3p}{(2\pi)^3}\frac{\mathrm{Tr}[P_E^{s_1}(\mathbf{p})\gamma_0\gamma_{\mu}P_E^{s_2}(\mathbf{p}+\mathbf{q})\gamma_0\gamma_\nu]}{i\omega_q-s_2\epsilon_0(\mathbf{p}+\mathbf{q})+s_1\epsilon_0(\mathbf{p})}[N^{s_2}(\mathbf{p}+\mathbf{q})(1-N^{s_1}(\mathbf{p}))-(1-N^{s_2}(\mathbf{p}+\mathbf{q}))N^{s_1}(\mathbf{p})],
\end{align}
\end{widetext}where $N^s(\mathbf{p})=\frac{1}{\exp[\beta(s\epsilon_0(\mathbf{p})-\mu)+1]}$,
$N^+(\mathbf{p})=N(\mathbf{p})$ is the Fermi distribution of the corresponding flavor and $\bar{N}(\mathbf{p})=\frac{1}{\exp[\beta(\epsilon_0(\mathbf{p})+\mu)]+1}=1-N^-(\mathbf{p})$
is the Fermi distribution for the corresponding anti-particle. The first line is
related to particle-hole creation and annihilation, the second and
third line to particle-antiparticle (and hole-antihole) production
and annihilation, and the last line to antiparticle-antiparticlehole
creation and annihilation. Note that, as expected, the first line
vanishes in the $T,\mu\rightarrow0$ limit and the last line vanishes
in the $T\rightarrow0$ limit. The second and third line contain infinite
vacuum ($T,\mu=0$) contributions, which need to be renormalized.
The gluon self-energy will be decomposed in terms of a matter part,
a renormalized vacuum and an infinite vacuum contribution according to
\[\Pi_{\mu\nu}(Q,T,\mu)=\Pi_{\mu\nu}^{\mathrm{mat}}+\Pi_{\mu\nu}^{\textrm{ren.vac.}}+\Pi_{\mu\nu}^{\textrm{inf.vac.}},\]
where the various functions have been defined as
\begin{align*}
\Pi_{\mu\nu}^{\textrm{mat}}&\equiv\Pi_{\mu\nu}(Q,T,\mu)-\frac{P_{\mu\nu}}{3}\Pi_{\lambda\lambda}(Q,0,0),\\
\Pi_{\mu\nu}^{\textrm{ren.vac.}}&\equiv\frac{P_{\mu\nu}}{3}\left[\Pi_{\lambda\lambda}(Q,0,0)-Q^2\left(\frac{\Pi_{\lambda\lambda}(Q,0,0)}{Q^2}\right)_{Q^2\rightarrow0}\right],\\
\Pi_{\mu\nu}^{\textrm{inf.vac.}}&\equiv\frac{P_{\mu\nu}}{3}Q^2\left[\frac{\Pi_{\lambda\lambda}(Q,0,0)}{Q^2}\right]_{Q^2\rightarrow0}.
\end{align*}
For the vacuum expressions it is possible to extract the projection matrix $P_{\mu\nu}$, since the polarization tensor is purely transverse and for $T,\mu=0$ there is no preferential frame such that the tensor is built up out of only two possible quantities $\delta_{\mu\nu}$ and $Q_\mu Q_\nu$. To include the dynamical properties of the vacuum expression a renormalization procedure is necessary. In the above the renormalized vacuum expression was obtained by extracting the infinite contributions from the vacuum polarization tensor at the renomalization point $Q^2=0$. The reason for extracting the factor $Q^2$ from the vacuum $\Pi_{\lambda\lambda}$ before setting $Q^2=0$ is because in the vacuum the dressed gluons are massless,
i.e., $\Pi_{\lambda\lambda}\propto Q^2$, c.f. Eq.~(\ref{eq:AP_renormalised_vacuum}). Therefore $Q^2=0$ is still a pole for the vacuum expression, at which the residue is $1$ after the renormalization. In conclusion, the following full renormalized polarization tensor will be used
\begin{equation}
\Pi_{\mu\nu}^{\textrm{ren}}=\Pi_{\mu\nu}-\Pi_{\mu\nu}^{\textrm{inf.vac.}}=\Pi_{\mu\nu}^{\mathrm{mat}}+\Pi_{\mu\nu}^{\textrm{ren.vac.}}.\label{eq:AP_renormalised_Pi}
\end{equation}

In the high-density limit ($\mu\gg m$), a good approximation is to
evaluate the matter part for the case of massless quarks. Thus for
massless quarks in the zero-temperature limit the matter parts of
$\Pi_{00}$ and $\Pi_{\mu\mu}$ can be calculated explicitly
\begin{widetext}
\begin{align}
\Pi_{00}^{\mathrm{mat}}(T=0)=&\frac{g^2}{2\pi^2}\left\{\frac{2}{3}\mu^2-\frac{1}{24}\sum_{s_1,s_2=\pm1}\left[q^2-\frac{is_1\omega_q}{2q}(Q^2+2q^2)\right]\ln\left[1-\frac{2s_1s_2\mu}{q-is_1\omega_q}\right]\right.\nonumber\\
&\qquad\qquad\qquad\qquad\qquad\qquad\qquad\left.+\sum_{s_2=\pm1}\left[s_2\frac{\mu(3Q^2-4\mu^2)}{24q}-\frac{\mu^2}{4}\frac{i\omega_q}{q}\right]\ln\left[\frac{1-\frac{2s_2\mu}{q-i\omega_q}}{1+\frac{2s_2\mu}{q+i\omega_q}}\right]\right\},\label{eq:AP_Pi_mat_00}\\
\Pi_{\mu\mu}^{\mathrm{mat}}(T=0)=&\frac{g^2}{\pi^2}\left[\frac{\mu^2}{2}-\frac{Q^2}{8q}\left\{\mu\sum_{s_1=\pm1}s_1\ln\left[\frac{1+\frac{2s_1\mu}{q-i\omega_q}}{1-\frac{2s_1\mu}{q+i\omega_q}}\right]+\frac{1}{2}\sum_{s_1,s_2=\pm1}(q-is_1\omega_q)\ln\left[1+\frac{2s_1s_2\mu}{q-is_1\omega_q}\right]\right\}\right].\label{eq:AP_Pi_mat_mumu}
\end{align}
\end{widetext}
These expressions give the same result as can be found
in Refs. \cite{Toimela:1985,Kapusta:1979,Kurkela:2010}.
Performing an analytic continuation to real time ($i\omega_q\rightarrow\omega+i0$),
the above expressions have branch cuts from $\max\left(0,q-2\mu\right)<|\omega|<q$
and $q<|\omega|<q+2\mu$, the origins of which can be found from Eq.~(\ref{PolarizationTensor})
and are due to particle-hole processes and finite-$T,\mu$ contributions
to particle-antiparticle production and annihilation, respectively.
Because Lorentz invariance is broken due to the presence of the gas,
$\Pi^{\textrm{mat}}$ is a function of $q_0=i(i\omega_q)$ and
$q$ separately since it can be a function of $q_\mu u_\mu$
and $q=\sqrt{Q^2-(q_\mu u_\mu)^2}$, where $u_\mu=(1,0,0,0)$
defines the rest frame of the system \cite{Kapusta:2006}.

The renormalized vacuum part $\Pi_{\mu\nu}^{\textrm{ren.vac.}}$ has
been found in Ref. \cite{Peskin:1995}
\begin{align}
P_{\mu\nu}&Q^2\frac{g^2}{4\pi^2}\int_0^1\mathrm{d}x\, x(1-x)\ln\left(\frac{m^2}{m^2+x(1-x)Q^2}\right)\nonumber\\
=&-\frac{1}{3}P_{\mu\nu}Q^2\frac{g^2}{4\pi^2}\left[\frac{1}{6}-\left(1-\frac{2m^2}{Q^2}\right)\right.\nonumber\\
&\times\left.\left(1-\sqrt{1+\frac{4m^2}{Q^2}}\mathrm{ArcCoth}\sqrt{1+\frac{4m^2}{Q^2}}\right)\right]\nonumber\\
\simeq&P_{\mu\nu}\frac{g^2Q^2\left[\frac{5}{3}+\ln\left(\frac{m^2}{Q^2}\right)\right]}{24\pi^2}.\label{eq:AP_renormalised_vacuum}
\end{align}
where it was assumed that the quarks have equal masses and in the
last line it was expanded for small masses. In real time this expression
has a branch cut for $|\omega|>\sqrt{q^2+4m^2}\simeq q$,
which is due to particle-antiparticle production and annihilation.

Using a Sommerfeld expansion temperature corrections to the gluon self-energy can be obtained. The Sommerfeld expansion can be summarized as
\begin{align}
N(\epsilon_p-\mu)=&\Theta(\mu-\epsilon_p)-\sum_{n=1}^{\infty}2(1-2^{1-2n})\zeta(2n)\nonumber\\
&\qquad\qquad\times\frac{\partial^{2n-1}}{\partial\epsilon_p^{2n-1}}\delta(\epsilon_p-\mu)T^{2n}.\label{eq:AP_Sommerfeld_Exp}
\end{align}
The derivation is analogous to that of the non-relativistic Sommerfeld
expansion \cite{Solyom:2008}. One has to keep in mind
that the above is not a complete expansion for small temperatures
since also the chemical potential depends on temperature. In this
manner the $T^{2}$ correction to the gluon self-energy is found to
be\begin{widetext}
\begin{align}
\Pi_{00}^{\mathrm{mat}}-\Pi_{00}^{\mathrm{mat}}(T=0)=&\frac{g^2}{2}\frac{1}{6}T^2\sum_{s=\pm1}\frac{i\omega_q+2s\mu}{2q}\ln\left[\frac{(i\omega_q-q)(i\omega_q+2s\mu+q)}{(i\omega_q+q)(i\omega_q+2s\mu-q)}\right],\nonumber\\
\Pi_{\mu\mu}^{\mathrm{mat}}-\Pi_{\mu\mu}^{\mathrm{mat}}(T=0)=&-\frac{g^2}{2}\frac{1}{6}T^2\frac{8\mu^2[3(i\omega_q)^2-(2\mu-q)(2\mu+q)]}{[(i\omega_q)^2-(2\mu-q)^2][(i\omega_q)^2-(2\mu+q)^2]}.\label{eq:AP_Pi_temperature_correction}
\end{align}
\end{widetext}Using Eqs.~(\ref{FG}-\ref{eq:AP_Pi_temperature_correction})
the behavior of $F$ and $G$ can be found in several limits. Keeping
the ratio $x=q/\omega$ fixed and expanding up to zeroth order in
$\omega$, the hard dense and hard thermal loop (HDL/HTL) expressions
are re-obtained \cite{Le Bellac:1996,Kapusta:2006}
\begin{align*}
\lim_{\omega\rightarrow0}&F(\omega,q=x\omega)\\=&2m_g^2\left.\left(1-\frac{1}{x^2}\right)\left[1+\frac{1}{2x}\ln\left(\frac{1-x}{1+x}\right)\right]\right|_{x=q/\omega},\\
\lim_{\omega\rightarrow0}&G(\omega,q=x\omega)\\=&m_g^2\left.\left[\frac{1}{x^2}-\frac{1}{2x}\left(1-\frac{1}{x^2}\right)\ln\left(\frac{1-x}{1+x}\right)\right]\right|_{x=q/\omega},
\end{align*}
where
\begin{equation}
m_g^2\equiv\frac{g^2}{4\pi^2}\left(\mu^2+\frac{1}{3}\pi^2T^2\right),\label{thermalmass}
\end{equation}
is the gluon thermal mass. The reason that the above limit returns the HDL and HTL expressions,
is due to the fact that the HDL/HTL approximation ($m=0$
and $\omega\ll q\ll\mu$) takes into account only the low-energy processes
around the Fermi surface, namely particle-hole processes. Some other
useful limits are
\begin{align*}
\lim_{q\rightarrow0}\lim_{\omega\rightarrow0}F&=2m_g^2\left(1+i\frac{\pi}{2}\frac{\omega}{q}\right),\\
\lim_{q\rightarrow0}\lim_{\omega\rightarrow0}G&=-2m_g^2i\frac{\pi}{2}\frac{\omega}{q},
\end{align*}
where it can be seen that in the static long-wavelength limit ($\omega,q\rightarrow0$) the electric gluons ($F$) are screened, while magnetic gluons ($G$) are dynamically screened. Furthermore, the large momenta and frequency behavior of the matter and vacuum parts are separately seen to be
\begin{align}
\lim_{q,\omega\rightarrow\infty}F^{\textrm{mat}}&=\frac{\mu^2\left(4m_g^2+\frac{5}{3}g^2T^2\right)}{3\mathrm{Q}^2},\nonumber\\
\lim_{q,\omega\rightarrow\infty}G^{\textrm{mat}}&=-\frac{\mu^2(4m_g^2+\frac{5}{3}g^2T^2)(q^2+\omega^2)}{3\mathrm{Q}^4},\nonumber\\
\lim_{\begin{array}{c}
{\scriptstyle q,\omega\rightarrow\infty}\\
{\scriptstyle T,\mu\rightarrow0}
\end{array}}F,G&=\frac{g^2\mathrm{Q}^2\left[\frac{5}{3}+\ln\left(\frac{m^2}{\mathrm{Q}^2}\right)\right]}{24\pi^2}.\label{eq:AP_FG_large_pomega_behavior}
\end{align}
The vacuum expressions are proportional to $\mathrm{Q}^2$ since in the vacuum
the gluons remain massless and thus $\mathrm{Q}^2=0$ is still a pole of
the propagator. The logarithmic $\mathrm{Q}^2$ dependence is usually absorbed
into the vertex and subsequently interpreted as the varying of the
coupling constant with the energy scale $\mathrm{Q}$.

\section{The thermodynamic potential of ideal gases}\label{app-thermpotential}

Consider first the well-known ideal Fermi gas term of a single species
\begin{align*}
-\frac{1}{V\beta}&\mathrm{Tr}\ln(-G_0^{-1})\\
&=-\frac{1}{V\beta}\sum_{\omega_n,p}\ln\det[(i\slashed{P}+m)]\\
&=-\frac{1}{\beta}\sum_{\omega_n}\int\frac{\mathrm{d}^3p}{(2\pi)^3}2\ln[-(i\omega_n+\mu_0)^2+\epsilon_0^2(\mathbf{p})].
\end{align*}
Using Eq.~(\ref{eq:AP_matsubara_fermions}) with $f(z)=\ln[-z^2+\epsilon_0^2(\mathbf{p})]$
and using the principle value logarithm with a branch cut on the negative real axis
\begin{align*}
\lim_{\eta\downarrow0}\Im[f(\omega+i\eta)]&=\lim_{\eta\downarrow0}\Im[\ln(-\omega^2+\epsilon_0^2(\mathbf{p})-2i\omega\eta)]\\
&=\pi\Theta[\omega^2-\epsilon_0^2(\mathbf{p})]\mathrm{sgn}(-\omega),
\end{align*}
the ideal Fermi gas contribution becomes
\begin{align*}
&-2\int\frac{\mathrm{d}^3p}{(2\pi)^3}\int_0^\infty[N(\omega-\mu_0)-(1-N(\omega+\mu_0))]\\
&\qquad\qquad\qquad\times\Theta[\omega^2-\epsilon_0^2(\mathbf{p})]\mathrm{d}\omega\\
=&-2\beta^{-1}\int\frac{\mathrm{d}^3p}{(2\pi)^3}\left\{\ln\left[1+e^{-\beta(\epsilon_0(\mathbf{p})-\mu_0)}\right]\right.\\
&\qquad\left.+\ln\left[1+e^{-\beta(\epsilon_0(\mathbf{p})+\mu_0)}\right]\right\}+2\int\frac{\mathrm{d}^3p}{(2\pi)^3}\int_{\epsilon_0^2(\mathbf{p})}^{\infty}d\omega.
\end{align*}
In the above it was assumed that the arc at infinity vanishes, which
is the case if the time-ordering in the path-integral is taken into
account properly by multiplying the integrand by $e^{-i\omega_n\eta}$
and taking the limit $\eta\downarrow0$. The infinite vacuum contribution
is clearly visible and should be subtracted, giving for massless fermions
\[\Omega_{\textrm{ideal-Fermi}}=-\left[\frac{7\pi^2T^4}{180}+\frac{T^2\mu_0^2}{6}+\frac{\mu_0^4}{12\pi^2}\right].\]

The ideal Bose gas follows from
\[\frac{1}{V\beta}\frac{1}{2}\mathrm{Tr}\left[\ln(D_0^{-1})-2\ln(\partial_\mu f_\mu)-\ln\frac{1}{\xi}\right],\]
which will lead to the same expression in both the Lorentz and Coulomb gauges. In the following, the Coulomb gauge is taken ($f_\mu=(0,\mathbf{\nabla})_\mu$),
\begin{align*}
\frac{1}{2V\beta}&\left(\mathrm{Tr}\ln[D_{0,\mu\nu}^{-1}(Q)]-2\mathrm{Tr}\ln q^2-\mathrm{Tr}\ln\frac{1}{\xi}\right)\\
&=\frac{1}{2V\beta}\sum_{\omega_n,q}\left(\ln\left[\frac{q^2}{\xi}q^2(Q^2)^2\right]-2\ln q^2-\ln\frac{1}{\xi}\right)\\
&=\frac{1}{V\beta}\sum_{\omega_n,q}\ln[-(i\omega_n)^2+q^2].
\end{align*}
The unphysical degrees of freedom clearly drop out due to the ghost
contribution $\partial_\mu f_\mu$. Using Eq.~(\ref{eq:AP_matsubara_bosons})
with $f(z)=\ln(-z^2+q^2)$, the ideal Bose
gas contribution is
\[\frac{2}{\beta}\int\frac{\mathrm{d}^3q}{(2\pi)^3}\ln[1-e^{-\beta q}]-\int\frac{\mathrm{d}^3q}{(2\pi)^3}\int_q^\infty\mathrm{d}\omega.\]
Again the vacuum term is clearly present and will be subtracted, giving
the Stefan-Boltzmann law
\[\Omega_\textrm{ideal-Bose}=-\frac{T^4}{45\pi^2}.\]

\section{The quark self-energy}\label{app-quark-selfenergy}

For completeness also the quark self-energy will be derived in the Lorentz gauge to lowest order.
\[
\Sigma(P)=\int\frac{\mathrm{d}^4Q}{\mathcal{V}}\mathrm{Tr}[-ig\gamma_\mu t^aG_0(Q)(-ig\gamma_\nu t^a)D_{\mu\nu}(P-Q)].
\]
Performing the Matsubara sum gives
\begin{widetext}
\begin{align}
\Sigma(i\omega_p,\mathbf{p})=&-g^2\frac{N_G}{2N_C}\int\frac{\mathrm{d}^3q}{(2\pi)^3}\frac{1}{2\epsilon^g(\mathbf{p}-\mathbf{q})}\nonumber\\
&\qquad\qquad\times\left\{\frac{\sum_s[\gamma_\mu u_s(\mathbf{q})\bar{u}_s(\mathbf{q})\gamma_\mu]}{i\omega_p+\mu-\epsilon_0(\mathbf{q})-\epsilon^g(\mathbf{p}-\mathbf{q})}[(1+N_B(\mathbf{p}-\mathbf{q}))(1-N(\mathbf{q}))+N_B(\mathbf{p}-\mathbf{q})N(\mathbf{q})]\right.\nonumber\\
&\qquad\qquad\quad\left.+\frac{\sum_s[\gamma_\mu u_s(\mathbf{q})\bar{u}_s(\mathbf{q})\gamma_\mu]}{i\omega_p+\mu-\epsilon_0(\mathbf{q})+\epsilon^g(\mathbf{p}-\mathbf{q})}[N_B(\mathbf{p}-\mathbf{q})(1-N(\mathbf{q}))+(1+N_B(\mathbf{p}-\mathbf{q}))N(\mathbf{q})]\right.\nonumber\\
&\qquad\qquad\quad\left.+\frac{\sum_s[\gamma_\mu v_s(\mathbf{q})\bar{v}_s(\mathbf{q})\gamma_\mu]}{i\omega_p+\mu+\epsilon_0(\mathbf{q})-\epsilon^g(\mathbf{p}-\mathbf{q})}[(1+N_B(\mathbf{p}-\mathbf{q}))\bar{N}(\mathbf{q})+N_B(\mathbf{p}-\mathbf{q})(1-\bar{N}(\mathbf{q}))]\right.\nonumber\\
&\qquad\qquad\quad\left.+\frac{\sum_s[\gamma_\mu v_s(\mathbf{q})\bar{v}_s(\mathbf{q})\gamma_\mu]}{i\omega_p+\mu+\epsilon_0(\mathbf{q})+\epsilon^g(\mathbf{p}-\mathbf{q})}[N_B(\mathbf{p}-\mathbf{q})\bar{N}(\mathbf{q})+(1+N_B(\mathbf{p}-\mathbf{q}))(1-\bar{N}(\mathbf{q}))]\right\},\nonumber\\
\Delta\Sigma(i\omega_p,\mathbf{p})=&-g^2\frac{N_G}{4N_C}\sum_{s_2,s_3=\pm1}\int\frac{\mathrm{d}^3q}{(2\pi)^3}\left[\frac{s_3}{\epsilon^g(\mathbf{p}-\mathbf{q})}\frac{\gamma_{\mu}P_E^{s_2}(\mathbf{q})\gamma_0\gamma_\mu}{i\omega_p+\mu-s_2\epsilon_0(\mathbf{q})-s_3\epsilon^g(\mathbf{p}-\mathbf{q})}\right]\nonumber\\
&\qquad\qquad\qquad\qquad\qquad\times[(1+N_B^{s_3}(\mathbf{p}-\mathbf{q}))(1-N^{s_2}(\mathbf{q}))+N_B^{s_3}(\mathbf{p}-\mathbf{q})N^{s_2}(\mathbf{q})-I_{s_2,s_3}].\label{eq:AP_Quark_Self_En}
\end{align}
\end{widetext}In the last line the indicator function $I_{s_2,s_3}=\delta_{s_2+}\delta_{s_3+}-\delta_{s_2-}\delta_{s_3-}$
is exactly the vacuum contribution such that $\Delta\Sigma$ is the
vacuum subtracted self-energy, i.e., $\Delta\Sigma\left(T=\mu=0\right)=0$.
To first order the self-energy for a single quark and antiquark can
be found by using the free quark energy projectors and helicity projectors
\[\Sigma_s^{s_1}(i\omega_p,\mathbf{p})=\mathrm{Tr}[\mathcal{P}^s(\mathbf{p})P_E^{s_1}\gamma_0\Sigma(i\omega_p,\mathbf{p})].\]
After projecting and evaluating the self-energy at $i\omega_p=s_1\epsilon_0(\mathbf{p})-\mu$ the function in the first square brackets in Eq.~(\ref{eq:AP_Quark_Self_En}) reduces
to $\mathcal{F}_{s_1,s_2,s_3}(\mathbf{p},\mathbf{q})$ defined in Eq.~(\ref{functionf}). In the zero-temperature limit the self-energy for a single quark then reduces to
\begin{align}
\lim_{T\rightarrow0}\Delta\Sigma^+(\epsilon_0(\mathbf{p})-\mu,\mathbf{p})=&\nonumber\\ \frac{g^2N_G}{4N_C}\sum_{s_3=\pm1}\int\frac{\mathrm{d}^3q}{(2\pi)^3}&\mathcal{F}_{+,+,s_3}(\mathbf{p},\mathbf{q})N^+(\mathbf{q}).\label{quarkselfenergy}
\end{align}
Note that the final result is independent of helicity $s$.

\end{document}